\documentclass[prb,reprint]{revtex4-1}
\usepackage{amsmath, amssymb}
\usepackage{amsfonts}
\usepackage{bbm} % to use mathbbm (e.g. identity 1)

\usepackage{graphicx}
\graphicspath{{images/}}

\usepackage{tikz}
\usetikzlibrary{shapes.geometric} % to draw regular polygons
\usetikzlibrary{positioning} % to use right=of a
\usetikzlibrary{calc} % to use relative coordinate
\tikzset{% global settings
    every picture/.style={line width=0.75pt},
    squ/.style = {regular polygon, regular polygon sides = 4, inner sep = 0, outer sep = 0, minimum size = 6.2mm, line width = 0.75pt},
    empty/.style = {inner sep = 0, outer sep = 0, minimum size = 0},
    tri/.style = {regular polygon, isosceles triangle, inner sep = 0, outer sep = 0, minimum width = 2.4mm, isosceles triangle apex angle = 50}
}

\usepackage{enumerate} % to enumerate items
\usepackage{enumitem} % to pause and resume enumeration

\usepackage{color} % to used colored texts
\usepackage{textcomp} % to use apostrophe
\usepackage{gensymb} % to use degree symbol
\definecolor{mylg}{gray}{0.55} % light gray for index markers

\usepackage{url}
\usepackage{hyperref}
\hypersetup{
    colorlinks=true,
    linkcolor = blue,
    filecolor = blue,
    citecolor = black,
    urlcolor  = black %cyan
}
\usepackage[nameinlink, capitalise]{cleveref}
\usepackage{verbatim} % to do multi-line comment
\usepackage{tabularx} % to use automatic line breaks in tables
\usepackage{amsmath} % to use alignat environment
\usepackage{lipsum}
\usepackage{bm} % to use bold greek letters

\usepackage[export]{adjustbox}

\newcommand{\imarker}[1]{{\footnotesize\color{mylg}#1}}
\def\basise{\vec{\mathrm{e}}}
\def\epsilonsmall{
    \includegraphics[scale=1.0,valign=c]{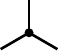}
}

\begin{document}

%%%%%%%%%%%%%%%%%%%%%%%%% Body %%%%%%%%%%%%%%%%%%%%%%%%%
%%%%%%% Title Block %%%%%%%
%\preprint{APS/123-QED}

\title{Boosting Vector Differential Calculus with the Graphical Notation}

\author{Joon-Hwi Kim}
\email{joonhwi.kim@gmail.com}
\affiliation{Department of Physics and Astronomy, Seoul National University, Seoul, South Korea}

\author{Maverick S. H. Oh}
\email{maverick.sh.oh@gmail.com}
\affiliation{Department of Physics and Photon Science, Gwangju Institute of Science and Technology, Gwangju, South Korea}

\author{Keun-Young Kim}
\email{fortoe@gist.ac.kr}
\homepage{https://phys.gist.ac.kr/gctp/}
\affiliation{Department of Physics and Photon Science, Gwangju Institute of Science and Technology, Gwangju, South Korea}

\date{\today}

\begin{abstract}
Learning vector calculus techniques is one of the major hurdles faced by physics undergraduates. However, beginners report various difficulties dealing with the index notation due to its bulkiness. Meanwhile, there have been graphical notations for tensor algebra that are intuitive and effective in calculations and can serve as a quick mnemonic for algebraic identities. Although they have been introduced and applied in vector algebra in the educational context, to the best of our knowledge, there have been no publications that employ the graphical notation to three-dimensional Euclidean vector \textit{calculus}, involving differentiation and integration of vector fields. Aiming for physics students and educators, we introduce such ``graphical vector calculus,'' demonstrate its pedagogical advantages, and provide enough exercises containing both purely mathematical identities and practical calculations in physics. The graphical notation can readily be utilized in the educational environment to not only lower the barriers in learning and practicing vector calculus
but also make students interested and self-motivated to manipulate the vector calculus syntax and heuristically comprehend the language of tensors by themselves.
\end{abstract}

\maketitle % title page is now complete
%%%%%%%%%%%%%%%%%%%%%%%%%%%

{
\section{Introduction\label{sec:Intro}}
As an essential tool in all fields of physics, vector calculus is one of the mathematical skills that physics undergraduates have to be acquainted with.
However, vector calculus with the index notation can be challenging to beginners due to its abstractness and bulkiness.
They report various difficulties: manipulating indices, getting lost and being ignorant about where to proceed toward during long calculations, memorizing the vector calculus identities, etc.
Meanwhile, there have been graphical languages for tensor algebra such as Penrose graphical notation, \cite{penrose1971negativedimensionaltensors} birdtracks,\cite{cvitanovic2008grouptheory, stedman2009diagramtechniques} or trace diagrams \cite{peterson2006tracediagram} that are intuitive and effective in calculations.
Although they can be readily applied to three-dimensional Euclidean vector calculus, publications covering vector calculus in a graphical notation remain absent in our best knowledge.
Previous works\cite{blinn2002quartic,stedman2009diagramtechniques,peterson2009unshacklingLA,peterson2006tracediagram,peterson2007notsocharacteristicequation,peterson2009onadiagrammaticproofofthecayley-hamiltontheorem,richter2009diagramstensorsandgeometricreasoning} only dealt with linear ``algebraic'' calculations and did not consider vector differential and integral ``calculus.''

In response to this, for physics learners and educators, we introduce the ``graphical vector calculus,'' advertise how easy and quick the graphical notation can derive vector calculus identities, and provide practical examples in the physics context.
Here, we consider differential calculus only; vector integral calculus might be covered in a following paper, as it also frequently appears in physics.
See the supplementary material\cite{SuppMat} for a brief discussion.
\par
Pedagogical advantages of the graphical notation are numerous.
First of all, it evidently resolves the aforementioned difficulties of a beginner. It serves as an intuitive language that is easy to acquire but does not lack any essential elements of vector calculus compared to the ordinary index notation.
In addition, students who are acquainted with the index notation would also benefit from learning the graphical notation. The graphical notation will increase their virtuosity in index gymnastics and promote them to develop concrete ideas of coordinate-free tensor algebra.
Lastly, the graphical notation of vector calculus serves as an excellent primer for graphical tools in modern physics such as perturbative diagrams in field theories as a conceptual precursor to Feynman diagrams. We anticipate that this ``user's manual'' of graphical vector calculus we provide will lower the barriers in learning and practicing vector calculus, as Feynman diagrams did in quantum field theory.

\section{Graphical Vector Algebra\label{sec:Algebra}}
\subsection{Motivation and Basic Rules}
We have two vectors, $\vec{A}$ and $\vec{B}$. We can make a scalar from these two by the dot product. In the ordinary index notation, we write $\vec{B}\cdot\vec{A} = B_iA_i$. Now, let us give some artistic touch to it.
\begin{equation}
    \includegraphics[scale=1.0,valign=c]{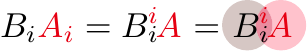}
\end{equation}
The ``$B$-atom'' and the ``$A$-atom'' are pairing their ``electrons'' (repeated index $i$) to form a ``covalent bond!'' Analogous to chemistry, depict a ``shared electron pair'' by a line connecting two ``atoms.''
\begin{equation}
    \label{eq:BdotA}
    \vec{B}\cdot\vec{A}=
    \adjustbox{valign=c}{\begin{tikzpicture}
        \node (B) at (0,0) [squ, draw] {$B$};
        \node (A) at (1,0) [squ, draw] {$A$};
        \draw (B) -- (A);
    \end{tikzpicture}}
\end{equation}
Vectors $\vec{A}$ and $\vec{B}$ are graphically represented as a box with a line attached to it.
The inner product is depicted by connecting the two lines of the two boxes.
Furthermore, an additional insight from this is that scalars will be graphically represented as objects with no ``external'' lines.  \adjustbox{valign=c,scale=0.8}{\begin{tikzpicture}
    \node (B) at (0,0) [squ, draw] {$B$};
    \node (A) at (1,0) [squ, draw] {$A$};
    \draw (B) -- (A);
\end{tikzpicture}}\,
only has an ``internal'' line; no lines are connected to the outside. It is isolated so that if the entire diagram is put inside a black box, no lines will poke out from it. In other words, scalars do not have free indices.
\begin{equation}
    {\renewcommand{\arraystretch}{1.5}
    \begin{array}{rl}
        \text{Scalars:}&
        \hspace{0.05em}f\hspace{0.078em}=\adjustbox{valign=c}{{\begin{tikzpicture}
            \node (f) at (0,0) [squ, draw] {$f$}; \end{tikzpicture}}}
        \\
        \text{Vectors:}&
        \vec{A}=\adjustbox{valign=c}{\begin{tikzpicture}
            \node (A) at (0,0) [squ, draw] {$A$}; \draw (A) -- (-1,0);
        \end{tikzpicture}}
    \end{array}}
\end{equation}
The basic observations here are summarized in \cref{tab:translationbwindexandgraphical}.

\newcolumntype{L}{>{\raggedright\arraybackslash}X}
\begin{table}[h]
    \begin{tabularx}{1.0\linewidth}{|L|L|}
        \hline
        \multicolumn{1}{|c|}{\textbf{Index Language}}
        & \multicolumn{1}{c|}{\textbf{Graphical Language}}
        \\ \hline
        An $n$-index quantity
        & A box with $n$ attached lines
        \\ \hline
        The name of a quantity
        & The character written inside the box
        \\ \hline
        Pairing (contracting) two indices
        & Connecting two ends of lines
        \\ \hline
        Free indices
        & External lines
        \\ \hline
        Contracted (dummy) indices
        & Internal lines
        \\ \hline
    \end{tabularx}
    \caption{Translation between the index language and the graphical language.}
    \label{tab:translationbwindexandgraphical}
\end{table}

Meanwhile, for scalar multiplication, addition, and subtraction, we do not introduce new notational rules to represent them but just borrow the ordinary notation; that is, they are denoted by juxtaposition and by ``$+$'' and ``$-$'' symbols.
\begin{equation}
    {\renewcommand{\arraystretch}{1.5}
        \begin{array}{rl}
        \text{Scalar multiplication:}&
        fg\hspace{0.21em}=\adjustbox{valign=c}{{\begin{tikzpicture}
            \node (f) at (0,0) [squ, draw] {$f$};
            \node (g) [right = 0.2cm of f] [squ, draw] {$g$};
            \end{tikzpicture}}}
        \\
        {}&
        f\vec{A}=\adjustbox{valign=c}{\begin{tikzpicture}
            \node (f) at (0,0) [squ, draw] {$f$};
            \node (A) [left = 0.2cm of f] [squ, draw] {$A$};
            \draw [black] (A) -- ($(A) + (-1,0)$);
        \end{tikzpicture}}
        \\
        \text{Addition\hspace{0.1em}/\hspace{0.1em}subtraction:}&
        f\hspace{0.05em}\pm\hspace{0.05em}g=\adjustbox{valign=c}{\begin{tikzpicture}
            \node (f) at (0,0) [squ, draw] {$f$};
            \node (plus) [right = 0.05cm of f] {$\pm$};
            \node (g) [right = 0.05cm of plus] [squ, draw] {$g$};
        \end{tikzpicture}}
        \\
        {}&
        \vec{A}\hspace{-0.125em}\pm\hspace{-0.175em}\vec{B}=\adjustbox{valign=c}{\begin{tikzpicture}
            \node (c) at (0.07,0) [empty] {};
            \node (A_tail) at (0,0) [empty] {};
            \node (A) at ($(A_tail)+(0.805, 0)+(c)$) [squ, draw] {$A$};
            \draw (A) -- (A_tail);
            \node (plus) at ($(A)+(0.435,0)$) [empty] {$\pm$};
            \node (B_tail) at ($(plus)+(0.195,0)$) [empty] {};
            \node (B) at ($(B_tail)+(0.810,0)+(c)$) [squ, draw] {$B$};
            \draw (B_tail) -- (B);
        \end{tikzpicture}}
    \end{array}}
\end{equation}
When two objects are juxtaposed, their relative position is irrelevant, such as
$\adjustbox{valign=c,scale=0.8}{{\begin{tikzpicture}
        \node (f) at (0,0) [squ, draw] {$f$};
        \node (g) at (0.55,0) [squ, draw] {$g$};
        \end{tikzpicture}}}
=
\adjustbox{valign=c,scale=0.8}{{\begin{tikzpicture}
        \node (f) at (0,0) [squ, draw] {$f$};
        \node (g) at (0.55,0.3) [squ, draw] {$g$};
        \end{tikzpicture}}}
=
\adjustbox{valign=c,scale=0.8}{{\begin{tikzpicture}
        \node (f) at (0,0) [squ, draw] {$f$};
        \node (g) at (-0.1,0.55) [squ, draw] {$g$};
        \end{tikzpicture}}}
=\cdots
$ etc.

However, it should be noted that in Eq. \ref{eq:BdotA}, $\vec{B}$ is depicted as a box with a line attached at its right side. It turns out that it is okay to not care about which side a line stems from a box for denoting vectors. A line can start from the left side, right side, upper side, lower side, or anywhere from the box, as if it freely ``dangles'' to be freely repositioned.
For example,
\begin{equation}
    \adjustbox{valign=c}{\begin{tikzpicture}
        \node (B) at (0,0) [squ, draw] {$B$};
        \node (A) at (1,0) [squ, draw] {$A$};
        \draw (B) -- (A);
    \end{tikzpicture}}
    =
    \adjustbox{valign=c}{\begin{tikzpicture}
        \node (A) at (0,0) [squ, draw] {$A$};
        \node (B) at (1,0) [squ, draw] {$B$};
        \draw (A) -- (B);
    \end{tikzpicture}}
    =
    \adjustbox{valign=c}{\begin{tikzpicture}[]
        \node (A) at (0,0) [squ, draw] {$A$};
        \node (B) at (0,-0.6) [squ, draw] {$B$};
        \draw [rounded corners = 0.3 cm] (A) -- ($(A)-(1,0)$) |- (B);
    \end{tikzpicture}}
    =
    \adjustbox{valign=c}{\begin{tikzpicture}[]
        \node (B) at (0,0) [squ, draw] {$B$};
        \node (A) at (0.6,0) [squ, draw] {$A$};
        \draw [rounded corners = 0.3 cm] (B) -- ($(B)-(0,-1)$) -| (A);
    \end{tikzpicture}}
    =\cdots,
    \label{eq:variousAdotBs1}
\end{equation}
and so on. It can be seen that an arbitrary rotation does not affect the value of a graphical equation.
Moreover, an arbitrary rearrangement of boxes also does not. For example, Eq. \ref{eq:variousAdotBs1} can be further deformed as the following.
\begin{equation}
    \includegraphics[scale=1.0,valign=c]{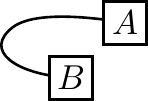}
    =
    \includegraphics[scale=1.0,valign=c]{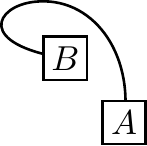}
    =
    \includegraphics[scale=1.0,valign=c]{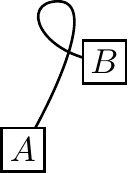}
    =
    \includegraphics[scale=1.0,valign=c]{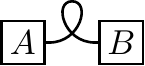}
    \label{eq:variousAdotBs2}
\end{equation}
So even if a diagram is drawn to look a little bit stiff, please remember that it is ``\textit{dancing}'' freely behind the scene!
Also, a line can freely pass under boxes, as you can see in the second equality in Eq. \ref{eq:variousAdotBs2}. In addition, intersections of lines have no significance; think of them just overpassing each other. When such intersections occur, we will always draw it in a manner that no ambiguity arises if one follows the ``law of good continuation.'' That is, ``\includegraphics[scale=1.0,valign=c]{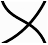}'' is an overlap of ``\includegraphics[scale=1.0,valign=c]{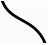}'' and ``\includegraphics[scale=1.0,valign=c]{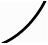},'' not ``\includegraphics[scale=1.0,valign=c]{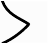}'' and ``\includegraphics[scale=1.0,valign=c]{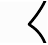}.''

\subsection{Meet the Kronecker Delta\label{sec:meettheKroneckerdelta}}
The diagram for $\vec{B}\cdot\vec{A}$ can be interpreted from a different perspective. The last diagram in  Eq. \ref{eq:variousAdotBs1} seems like two vectors
\adjustbox{scale=0.8, valign=c}{
    \adjustbox{valign=c}{\begin{tikzpicture}
        \node (A) at (0,0) [squ, draw] {$B$}; \draw (A) -- (0,0.65);
    \end{tikzpicture}}
} and \adjustbox{scale=0.8, valign=c}{
    \adjustbox{valign=c}{\begin{tikzpicture}
        \node (A) at (0,0) [squ, draw] {$A$}; \draw (A) -- (0,0.65);
    \end{tikzpicture}}
} are ``plugged into'' a \includegraphics[valign=c]{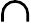}-shaped object.
\begin{equation}
    \includegraphics[scale=1.0,valign=c]{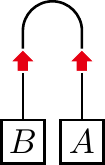}
\end{equation}
Then, what does the \includegraphics[valign=c]{figs/cap.pdf}-shaped object represent? It is a ``machine''\cite{Note1} that takes two vectors as input and gives a scalar; it is the inner product ``$\,\cdot\,$,'' or in the index notation, ``$\delta_{ij}$.'' Plugging lines into the machine corresponds to contraction of indices.
\begin{alignat}{3}
    B_i \delta_{ij} A_j
    &=
    \adjustbox{valign=c}{\begin{tikzpicture}[]
        \node (B) at (0,0) [squ, draw] {$B$};
        \node (A) at (0.6,0) [squ, draw] {$A$};
        \draw [rounded corners = 0.3 cm] (B) -- ($(B)-(0,-1)$) -| (A);
    \end{tikzpicture}}
    &&=
    \adjustbox{valign=c}{\begin{tikzpicture}
        \node (B) at (0,0) [squ, draw] {$B$};
        \node (A) at (1,0) [squ, draw] {$A$};
        \draw (B) -- (A);
    \end{tikzpicture}}
    &&= \cdots\,;
    \\
    \delta_{ij}
    &=\,\,
    \adjustbox{valign=c}{\begin{tikzpicture}[]
        \node (B) at (0,0) [inner sep = 2pt] {\imarker{$i$}};
        \node (A) at (0.6,0) [inner sep = 2pt] {\imarker{$j$}};
        \draw [rounded corners = 0.3 cm] (B) -- ($(B)-(0,-1)$) -| (A);
    \end{tikzpicture}}
    &&=\,\,
    \adjustbox{valign=c}{\begin{tikzpicture}
        \node (B) at (0,0) [inner sep = 2pt] {\imarker{$i$}};
        \node (A) at (1,0) [inner sep = 2pt] {\imarker{$j$}};
        \draw (B) -- (A);
    \end{tikzpicture}}
    &&=\cdots
    \,.
\end{alignat}
In the second line, we turned on the ``index markers'' to avoid confusion that which terminal of the line corresponds to the index $i$ and $j$, respectively.

A comment should be made about the symmetry of the Kronnecker delta. The fact that $\delta_{ij}=\delta_{ji}$ is \textit{already reflected in the design} of our graphical notation, that is the appearance of $\delta_{ij}$ with the \textit{dancing rule of equivalent diagrams}. In the graphical notation, $\delta_{ij}$ is an undirected line, so that there is no way to distinguish its ``left'' and ``right'' terminals. For instance, see the first equality of Eq. \ref{eq:variousAdotBs1}. If you want to write this symmetry condition without ``test vectors'' plugged in, observe the second form of $\vec{B}\cdot\vec{A}$ in Eq. \ref{eq:variousAdotBs1} and the last form in Eq. \ref{eq:variousAdotBs2}. It can be seen that
\begin{equation}
    \includegraphics[scale=1.0,valign=c]{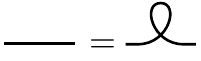}\,.
    \label{eq:deltatwist}
\end{equation}
Turning on the index markers,\cite{Note2}
\begin{equation}
    \includegraphics[scale=1.0,valign=c]{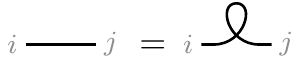}\,,
\end{equation}
or giving one more touch,
\begin{equation}
    \includegraphics[scale=1.0,valign=c]{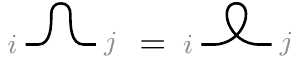}\,.
    \label{eq:deltatwistimon}
\end{equation}
The left hand side assigns $i$ to the left terminal of the \includegraphics[valign=c]{figs/cap.pdf}-shaped and $j$ to the right terminal; the right hand side assigns $i$ to the right terminal and $j$ to the left terminal. Just pretend for a moment that the index assigned to the left terminal should be placed first when reading the \includegraphics[valign=c]{figs/cap.pdf}-shaped in Eq. \ref{eq:deltatwistimon} in the index notation; then, we have $\delta_{ij}=\delta_{ji}$.

\subsection{Meet the Cross Product Machine\label{sec:meetthecrosspoduct}}
Now, move on to the next important structure, the cross product. The cross product is a machine that takes two vectors as a input and gives a vector. Hence, two lines are needed for input and one line for output.
\begin{equation}
    \vec{A}\hspace{-0.03em}\times\hspace{-0.03em}\vec{B}
    \hspace{-0.08em}=\hspace{-0.08em}\hspace{-0.01em}
    \adjustbox{valign=c}{
    \begin{tikzpicture}[]
        \node (O) at (0,0) [empty] {};
        \node (i) at ( 90:0.6) [empty] {};
        \node (j) at (-{0.3*sqrt(3)},-0.6) [squ,draw] {$A$};
        \node (k) at (+{0.3*sqrt(3)},-0.6) [squ,draw] {$B$};
        \fill [fill = black] (O) circle (1.3pt);
        \draw (i) -- (O);
        \draw (j) .. controls ($(210:0.55)$).. (O) .. controls ($(-30:0.55)$).. (k);
        \node (phantom) at ($(0,-1.2)+(0,-0.31)$) [empty] {};
    \end{tikzpicture}
    }
    \hspace{-0.01em}\hspace{-0.08em}=\hspace{-0.08em}\hspace{-0.76em}
    \adjustbox{valign=c}{
    \begin{tikzpicture}[]
        \node (O) at (0,0) [empty] {};
        \node (i) at ( 90:0.6) [empty] {};
        \node (j) at (0.4,-1.2) [squ,draw] {$A$};
        \node (k) at (+{0.3*sqrt(3)},-0.6) [squ,draw] {$B$};
        \fill [fill = black] (O) circle (1.3pt);
        \draw (i) to (O);
        \draw (j) to [out=120, in=210, looseness=1.5] (O) .. controls ($(-30:0.55)$).. (k);
        \node (phantom) at ($(0,-1.2)+(0,-0.31)$) [empty] {};
    \end{tikzpicture}
    }
    \hspace{-0.01em}\hspace{-0.08em}=\hspace{-0.08em}\hspace{-0.01em}
    \adjustbox{valign=c}{
    \begin{tikzpicture}[]
        \node (O) at (0,0) [empty] {};
        \node (i) at ( 90:0.6) [empty] {};
        \node (j) at (-{0.3*sqrt(3)},-0.6) [squ,draw] {$B$};
        \node (k) at (+{0.3*sqrt(3)},-0.6) [squ,draw] {$A$};
        \fill [fill = black] (O) circle (1.3pt);
        \draw (i) -- (O);
        \draw (k) .. controls ($(210:0.4)$).. (O) .. controls ($(-30:0.4)$).. (j);
        \node (phantom) at ($(0,-1.2)+(0,-0.31)$) [empty] {};
    \end{tikzpicture}
    }
    \hspace{-0.01em}\hspace{-0.08em}=\hspace{-0.08em}
    \cdots
    \label{eq:crossproduct}
\end{equation}
Please do not forget the diagrams are dancing and  Eq. \ref{eq:crossproduct} is showing just three snapshots. There are infinitude of possible configurations that $\vec{A}\times\vec{B}$ can be drawn.
Also, note that the third diagram is read as $\vec{A} \times \vec{B}$ as well as the first one.
The lines attached to the cross product machine (\epsilonsmall) should be read counterclockwise from the core (the small dot) of the machine: \includegraphics[valign=c,scale=1]{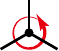}.
The left and right arms of the cross product machine is connected to $\vec{A}$ and $\vec{B}$ respectively in both the first and third diagrams in Eq. \ref{eq:crossproduct}, so they are equivalent.
Continuous deformations do not affect the value of a diagram.

However, how about discontinuous deformations? In case of the inner product, yanking a twist, a discontinuous deformation that yields a cusp during the process, did not affect the value because the inner product is symmetric. In case of the cross product, it is antisymmetric so that $\vec{A}\times\vec{B}=-\vec{B}\times\vec{A}$; therefore, when the two arms of the first diagram in Eq. \ref{eq:crossproduct} are swapped---which is the third diagram---and yanked, a minus sign pops out, as depicted in \cref{fig:yankclank}. Associating a kinesthetic imagery that the lines of the cross product machine are elastic but particularly stiff near the core might be helpful to intuitively remember this. Do not forget the minus sign. Yanking a twist is a discontinuous ``clank'' process.

Note that in case of a general object (tensor), the value after swap-then-yanking its two arms is by no means related to the original value, unless it bears symmetry or antisymmetry with respect to permutation of the two indices.

\begin{figure}[t]
    \centering
    \includegraphics[width=1.0\linewidth]{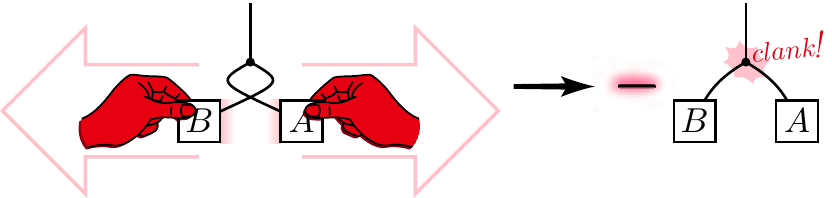}
    \caption{A minus sign pops out with a ``clank!'' sound when you swap-then-yank the two arms of a cross product machine. The plaintext equation corresponding to this action is ``$\vec{A}\times\vec{B}=-\vec{B}\times\vec{A}$.''}
    \label{fig:yankclank}
\end{figure}

\subsection{Triple Products\label{sec:tripleproducts}}
Having introduced the graphical notation for the cross product, let us now graphically express triple product identities. First, a scalar triple product $\vec{C}\cdot\big(\vec{A}\times\vec{B}\big)$ can be drawn by connecting the free terminals of
\adjustbox{scale=0.8, valign=c}{
    \adjustbox{valign=c}{\begin{tikzpicture}
        \node (A) at (0,0) [squ, draw] {$C$}; \draw (A) -- (0,0.65);
    \end{tikzpicture}}
}
and Eq. \ref{eq:crossproduct}:
\begin{equation}
    \label{eq:ABC}
    \adjustbox{valign=c}{
    \begin{tikzpicture}[]
        \node (C) at (-1.2,-0.3) [squ, draw] {$C$};
        \node (O) at (0,0) [empty] {};
        \node (i) at (0,0.6) [empty] {};
        \node (j) at (210:0.6) [squ,draw] {$A$};
        \node (k) at (-30:0.6) [squ,draw] {$B$};
        \fill [fill = black] (O) circle (1.3pt);
        \draw (i) -- (O);
        \draw (j) -- (O) -- (k);
        \node (m) at (-0.6, 1.2) [empty] {};
        \draw [rounded corners = 0.6cm] (C) |-(m)-| (i);
    \end{tikzpicture}
    }
    =
    \adjustbox{valign=c}{
    \begin{tikzpicture}[]
        \node (O) at (0,0) [empty] {};
        \node (i) at ( 90:0.6) [squ,draw] {$C$};
        \node (j) at (210:0.6) [squ,draw] {$A$};
        \node (k) at (-30:0.6) [squ,draw] {$B$};
        \fill [fill = black] (O) circle (1.3pt);
        \draw (i) -- (O);
        \draw (j) -- (O) -- (k);
    \end{tikzpicture}
    }
    \,.
\end{equation}
The cyclic symmetry of the scalar triple product is \textit{already reflected in its graphical design}: it looks the same under threefold rotation.
\begin{equation}
    {\renewcommand{\arraystretch}{1.5}
    \begin{array}{ccc}
        \adjustbox{valign=c}{
        \begin{tikzpicture}[]
            \node (O) at (0,0) [empty] {};
            \node (i) at ( 90:0.6) [squ,draw] {$C$};
            \node (j) at (210:0.6) [squ,draw] {$A$};
            \node (k) at (-30:0.6) [squ,draw] {$B$};
            \fill [fill = black] (O) circle (1.3pt);
            \draw (i) -- (O);
            \draw (j) -- (O) -- (k);
        \end{tikzpicture}
        }
        &=
        \adjustbox{valign=c}{
        \begin{tikzpicture}[]
            \node (O) at (0,0) [empty] {};
            \node (i) at ( 90:0.6) [squ,draw] {$A$};
            \node (j) at (210:0.6) [squ,draw] {$B$};
            \node (k) at (-30:0.6) [squ,draw] {$C$};
            \fill [fill = black] (O) circle (1.3pt);
            \draw (i) -- (O);
            \draw (j) -- (O) -- (k);
        \end{tikzpicture}
        }
        &=
        \adjustbox{valign=c}{
        \begin{tikzpicture}[]
            \node (O) at (0,0) [empty] {};
            \node (i) at ( 90:0.6) [squ,draw] {$B$};
            \node (j) at (210:0.6) [squ,draw] {$C$};
            \node (k) at (-30:0.6) [squ,draw] {$A$};
            \fill [fill = black] (O) circle (1.3pt);
            \draw (i) -- (O);
            \draw (j) -- (O) -- (k);
        \end{tikzpicture}
        }
        \\
        {}&{\hspace{1.08em}\updownarrow}&{}
        \\
        \vec{C}\cdot\big(\vec{A}\times\vec{B}\big)
        &=
        \vec{A}\cdot\big(\vec{B}\times\vec{C}\big)
        &=
        \vec{B}\cdot\big(\vec{C}\times\vec{A}\big)
        \\
    \end{array}}
\end{equation}
This is \textit{the economy of graphical notations}: redundant plaintext expressions are brought to the same or at least manifestly equivalent diagram.

As a side note, imagine what would it mean if the cross product machine is naked, while it is fully dressed in Eq. \ref{eq:ABC}, which is $\epsilon_{ijk}C_iA_jB_k$ in the index notation. As some readers might already noticed, another name for the cross product machine is the Levi-Civita symbol, $\epsilon_{ijk}$. It is a three-terminal machine (three-index tensor), and antisymmetric in every pair of its arms (indices).
\begin{equation}
    \epsilon_{ijk}\, =
    \hspace{-0.16em}
    \adjustbox{valign=c}{
    \begin{tikzpicture}[]
        \node (O) at (0,0) [empty] {};
        \node (i) at ( 90:0.6) [inner sep = 2pt, outer sep = 0pt] {\imarker{$k$}};
        \node (j) at (210:0.6) [inner sep = 2pt, outer sep = 0pt] {\imarker{$i$}};
        \node (k) at (-30:0.6) [inner sep = 2pt, outer sep = 0pt] {\imarker{$j$}};
        \fill [fill = black] (O) circle (1.3pt);
        \draw (i) -- (O);
        \draw (j) -- (O) -- (k);
    \end{tikzpicture}
    }
\end{equation}

Next is the vector triple product. The BAC-CAB formula translates into the graphical language as the following.
\begin{equation}
    {\renewcommand{\arraystretch}{1.5}
    \begin{array}{ccccc}
        \adjustbox{valign=c}{
        \begin{tikzpicture}[]
            \node (O) at (0,0) [empty] {};
            \node (j) at (-{0.3*sqrt(3)},-0.6) [squ,draw] {$B$};
            \node (k) at (+{0.3*sqrt(3)},-0.6) [squ,draw] {$C$};
            \fill [fill = black] (O) circle (1.3pt);
            \draw (j) .. controls ($(210:0.55)$).. (O) .. controls ($(-30:0.55)$).. (k);
            \node (A) at (-{0.6*sqrt(3)},0) [squ,draw] {$A$};
            \node (Q) at (-{0.3*sqrt(3)},0.6) [empty] {};
            \node (e) at (-{0.3*sqrt(3)},1.2) [empty] {};
            \fill [fill = black] (Q) circle (1.3pt);
            \draw (e) -- (Q);
            \draw (A) .. controls ($(Q)+(210:0.55)$).. (Q) to[out=-30,in=90] (O);
        \end{tikzpicture}
        \hspace{-1.0em}
        }
        &
        {=}
        &
        {
        \includegraphics[scale=1.0,valign=c]{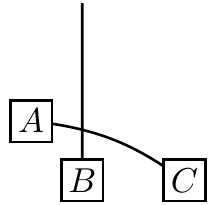}
        \hspace{-0.5em}
        }
        &
        {-}
        &
        {
        \includegraphics[scale=1.0,valign=c]{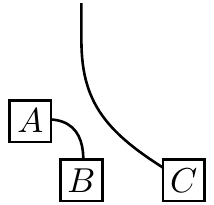}
        \hspace{-0.5em}
        }
        \\
        {}&{}&{\updownarrow}&{}&{}
        \\
        {\vec{A}\times\big(\vec{B}\times\vec{C})
        \hspace{-0.8em}
        }
        &
        {=}
        &
        {\vec{B} (\vec{A}\cdot\vec{C})
        \hspace{-0.5em}
        }
        &
        {-}
        &
        {\vec{C}(\vec{A}\cdot\vec{B})
        \hspace{-0.5em}
        }
        \\
    \end{array}}
\end{equation}
This holds for arbitrary $\vec{A}$, $\vec{B}$, and $\vec{C}$; thus, one can extract the ``bones'' only:
\begin{equation}
    \adjustbox{valign=c}{
    \begin{tikzpicture}
        \node (O) at (0,0) [empty] {};
        \node (alpha) at (0,0.2) [empty] {};
        \node (beta) at (0,-0.2) [empty] {};
        \node (xi) at ($(alpha)+(30:0.4)$) [empty] {};
        \node (xj) at ($(alpha)+(150:0.4)$) [empty] {};
        \node (xl) at ($(beta)+(-150:0.4)$) [empty] {};
        \node (xm) at ($(beta)+(-30:0.4)$) [empty] {};
        \node (i) at ($(xi)+(0,0.2)$) [empty] {};
        \node (j) at ($(xj)+(0,0.2)$) [empty] {};
        \node (l) at ($(xl)+(0,-0.2)$) [empty] {};
        \node (m) at ($(xm)+(0,-0.2)$) [empty] {};
        \fill (alpha) circle (1.3pt);
        \fill (beta) circle (1.3pt);
        \draw (alpha) -- (O) node [anchor = west] {} -- (beta);
        \draw (i) ..controls(xi).. (alpha) ..controls(xj).. (j);
        \draw (l) ..controls(xl).. (beta) ..controls(xm).. (m);
    \end{tikzpicture}}
    \,\,=\,\,
    \adjustbox{valign=c}{
    \begin{tikzpicture}
        \node (O) at (0,0) [empty] {};
        \node (alpha) at (0,0.2) [empty] {};
        \node (beta) at (0,-0.2) [empty] {};
        \node (xi) at ($(alpha)+(30:0.4)$) [empty] {};
        \node (xj) at ($(alpha)+(150:0.4)$) [empty] {};
        \node (xl) at ($(beta)+(-150:0.4)$) [empty] {};
        \node (xm) at ($(beta)+(-30:0.4)$) [empty] {};
        \node (i) at ($(xi)+(0,0.2)$) [empty] {};
        \node (j) at ($(xj)+(0,0.2)$) [empty] {};
        \node (l) at ($(xl)+(0,-0.2)$) [empty] {};
        \node (m) at ($(xm)+(0,-0.2)$) [empty] {};
        \draw (i) ..controls ($(xi)-(0,0.2)$) and ($(xl)+(0,0.2)$) .. (l);
        \draw (j) ..controls ($(xj)-(0,0.2)$) and ($(xm)+(0,0.2)$) .. (m);
    \end{tikzpicture}}
    \,\,-\,\,
    \adjustbox{valign=c}{
    \begin{tikzpicture}
        \node (O) at (0,0) [empty] {};
        \node (alpha) at (0,0.2) [empty] {};
        \node (beta) at (0,-0.2) [empty] {};
        \node (xi) at ($(alpha)+(30:0.4)$) [empty] {};
        \node (xj) at ($(alpha)+(150:0.4)$) [empty] {};
        \node (xl) at ($(beta)+(-150:0.4)$) [empty] {};
        \node (xm) at ($(beta)+(-30:0.4)$) [empty] {};
        \node (i) at ($(xi)+(0,0.2)$) [empty] {};
        \node (j) at ($(xj)+(0,0.2)$) [empty] {};
        \node (l) at ($(xl)+(0,-0.2)$) [empty] {};
        \node (m) at ($(xm)+(0,-0.2)$) [empty] {};
        \draw (i) -- (m);
        \draw (j) -- (l);
    \end{tikzpicture}}\,\,\,.
    \label{eq:theidentity}
\end{equation}
Until now, all graphical equations followed from defining rules of graphical representation. However, Eq. \ref{eq:theidentity} is the first---and indeed the only---nontrivial formula relating cross product machines and Kronecker deltas. This is the most important identity that serves as a basic ``syntax'' of our calculations.

Equation \ref{eq:theidentity} is by no means ``new.'' With the index markers, it turns out that it is the well-known formula about contracted two $\epsilon_{ijk}$'s.
\begin{alignat}{3}
    &
    \adjustbox{valign=c}{
    \begin{tikzpicture}
        \node (O) at (0,0) [empty] {};
        \node (k) at (O) [empty] {\imarker{\hspace{1.1em}$k$}};
        \node (alpha) at (0,0.2) [empty] {};
        \node (beta) at (0,-0.2) [empty] {};
        \node (xi) at ($(alpha)+(30:0.4)$) [empty] {};
        \node (xj) at ($(alpha)+(150:0.4)$) [empty] {};
        \node (xl) at ($(beta)+(-150:0.4)$) [empty] {};
        \node (xm) at ($(beta)+(-30:0.4)$) [empty] {};
        \node (i) at ($(xi)+(0,0.2)+(0,0.15301)$) [inner sep = 1.04pt] {\imarker{$i$}};%0.15301->고유명사처럼 활용
        \node (j) at ($(xj)+(0,0.2)+(0,0.15301)$) [inner sep = 1.04pt] {\imarker{$j$}};
        \node (l) at ($(xl)+(0,-0.2)-(0,0.15301)$) [inner sep = 1.04pt] {\imarker{$l$}};
        \node (m) at ($(xm)+(0,-0.2)-(0,0.15301)$) [inner sep = 1.04pt] {\imarker{$m$}};
        \fill (alpha) circle (1.3pt);
        \fill (beta) circle (1.3pt);
        \draw (alpha) -- (O) node [anchor = west] {} -- (beta);
        \draw (i) ..controls(xi).. (alpha) ..controls(xj).. (j);
        \draw (l) ..controls(xl).. (beta) ..controls(xm).. (m);
    \end{tikzpicture}}
    &&=\,
    \adjustbox{valign=c}{
    \begin{tikzpicture}
        \node (O) at (0,0) [empty] {};
        \node (alpha) at (0,0.2) [empty] {};
        \node (beta) at (0,-0.2) [empty] {};
        \node (xi) at ($(alpha)+(30:0.4)$) [empty] {};
        \node (xj) at ($(alpha)+(150:0.4)$) [empty] {};
        \node (xl) at ($(beta)+(-150:0.4)$) [empty] {};
        \node (xm) at ($(beta)+(-30:0.4)$) [empty] {};
        \node (i) at ($(xi)+(0,0.2)+(0,0.15301)$) [inner sep = 1.04pt] {\imarker{$i$}};
        \node (j) at ($(xj)+(0,0.2)+(0,0.15301)$) [inner sep = 1.04pt] {\imarker{$j$}};
        \node (l) at ($(xl)+(0,-0.2)-(0,0.15301)$) [inner sep = 1.04pt] {\imarker{$l$}};
        \node (m) at ($(xm)+(0,-0.2)-(0,0.15301)$) [inner sep = 1.04pt] {\imarker{$m$}};
        \draw (i) ..controls ($(xi)-(0,0.2)$) and ($(xl)+(0,0.2)$) .. (l);
        \draw (j) ..controls ($(xj)-(0,0.2)$) and ($(xm)+(0,0.2)$) .. (m);
    \end{tikzpicture}}
    &&-\,\,
    \adjustbox{valign=c}{
    \begin{tikzpicture}
        \node (O) at (0,0) [empty] {};
        \node (alpha) at (0,0.2) [empty] {};
        \node (beta) at (0,-0.2) [empty] {};
        \node (xi) at ($(alpha)+(30:0.4)$) [empty] {};
        \node (xj) at ($(alpha)+(150:0.4)$) [empty] {};
        \node (xl) at ($(beta)+(-150:0.4)$) [empty] {};
        \node (xm) at ($(beta)+(-30:0.4)$) [empty] {};
        \node (i) at ($(xi)+(0,0.2)+(0,0.15301)$) [inner sep = 1.04pt] {\imarker{$i$}};
        \node (j) at ($(xj)+(0,0.2)+(0,0.15301)$) [inner sep = 1.04pt] {\imarker{$j$}};
        \node (l) at ($(xl)+(0,-0.2)-(0,0.15301)$) [inner sep = 1.04pt] {\imarker{$l$}};
        \node (m) at ($(xm)+(0,-0.2)-(0,0.15301)$) [inner sep = 1.04pt] {\imarker{$m$}};
        \draw (i) -- (m);
        \draw (j) -- (l);
    \end{tikzpicture}}
    \\
    &{}&&{\hspace{2.99em}\updownarrow}&&{} \nonumber
    \\
    &
    {\epsilon_{ijk}\epsilon_{klm}}
    &&=
    {\delta_{jm}\delta_{il}}
    &&-
    {\delta_{jl}\delta_{im}}
    \label{eq:theidentityinindexnotaion}
\end{alignat}
However, the graphical way has multiple appealing points. First, it naturally serves as a quick visual mnemonic for Eq. \ref{eq:theidentityinindexnotaion}. Also, in practical circumstances, the graphical form avoids the bulkiness of dummy indices and significantly simplifies the procedure of index replacement by $\delta_{ij}$'s. One does not have to say ``$i$ to $l$, $j$ to $m$'' over and over in one's mind organizing the expanded terms. This makes a greater difference in calculation time as the equation involves more operations and dummy indices (proof of the Jacobi identity,\cite{SuppMat} for example). On the other hand, classification of vector algebraic identities is immediate if they are written in the graphical notation, because it shows the (tensorial) structure of equations explicitly.
One can recognize identical structures within a single glance, as comprehension of visuals is much faster than that of texts. Some may take a critical stance to this, because mere counting of the symbols ``$\times$'' and ``\,$\cdot$\,'' would also reveal the structure of equations, albeit for simple cases. However, with the graphical notation, generating different identities of the same structure is also straightforward; it is accomplished by just attaching ``flesh pieces'' (vectors or arbitrary multi-terminal objects\cite{Note3}) to the ``bone.'' For instance, one can easily write down the equations equivalent to the BAC-CAB rule or the Jacobi identity.\cite{SuppMat} Knowing what fundamental rules that identities are rooted in with being able to generate equivalent identities will effectively promote concrete understandings of the structure of vector algebra.

\section{Graphical Vector Calculus\label{sec:Calculus}}
Now is the time for graphical vector ``calculus.''
Here, we are considering not just scalars and vectors, but ``scalar fields'' $f(\vec{r}\hspace{0.1em})$, $g(\vec{r}\hspace{0.1em})$, $\cdots$ and ``vector fields'' $\vec{A}(\vec{r}\hspace{0.1em})$, $\vec{B}(\vec{r}\hspace{0.1em})$, $\cdots$; they depend on spatial coordinates, or equivalently, the position vector $\vec{r}$. In this section, ``$(\vec{r}\hspace{0.1em})$'' is omitted unless there is an ambiguity whether it depends on $\vec{r}$ or not.

\subsection{The Basics}
The first mission would be graphically representing $\nabla = \basise_i\frac{\partial}{\partial x_i}:=\basise_i\partial_{i}$, where $\basise_i$ and $x_i$ are the $i^{\text{th}}$ Cartesian basis vector and coordinate, respectively. $\nabla$ is a ``vector'' (that is, it carries an index), but also a differential operator at the same time. Therefore, to accomplish the mission, a notation that has one terminal and is capable of representing the Leibniz property (the product rule of derivatives) should be devised. The later can be achieved by an empty circle, which reminds of a balloon. Things inside the balloon are subjected to differentiation. The balloon ``eats'' $fg$ by first biting $f$ only then $g$ only:
$
\adjustbox{valign=c, scale=0.8}{
\begin{tikzpicture}[]
    \node (f) [squ, draw] at (0,0) {$f$};
    \node (g) [right = 0.1cm of f] [squ, draw] {$g$};
    \node (dif) [empty] at ($(f)!0.5!(g)$) {\includegraphics[scale=1.0]{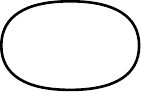}};
\end{tikzpicture}
}
\hspace{-0.25em}
=
\hspace{-0.25em}
\adjustbox{valign=c, scale=0.8}{
\begin{tikzpicture}[]
    \node (f) [squ, draw] at (0,0) {$f$};
    \node (g) [right = 0.4cm of f] [squ, draw] {$g$};    \node (dif) [circle, draw, minimum size = 9mm] at (f) {};
\end{tikzpicture}
}
\hspace{-0.06em}
+
\hspace{-0.06em}
\adjustbox{valign=c, scale=0.8}{
\begin{tikzpicture}[]
    \node (f) [squ, draw] at (0,0) {$f$};
    \node (g) [right = 0.4cm of f] [squ, draw] {$g$};
    \node (dif) [circle, draw, minimum size = 9mm] at (g) {};
\end{tikzpicture}
}
$ $\hspace{-0.1em}\leftrightarrow (fg)'=f'g+fg'$. To ``vectorize'' this, we simply attach a single tail to it.
\begin{align}
    \label{eq:Leibnizrule}
    {\renewcommand{\arraystretch}{1.5}
    \begin{array}{ccccc}
        \adjustbox{valign=c, scale=1}{
        \begin{tikzpicture}[]
            \node (f) [squ, draw] at (0,0) {$f$};
            \node (g) [right = 0.1cm of f] [squ, draw] {$g$};
            \node (dif) [empty] at ($(f)!0.5!(g)$) {\includegraphics[scale=1.0]{figs/CCSqBalloon9x14mm.pdf}};
            \node (tail) at ($(dif)+(0,-1.4)$) [inner sep = 2pt] {\imarker{$i$}};
            \draw (dif)--(tail);
        \end{tikzpicture}
        }
        &
        =
        &
        \adjustbox{valign=c, scale=1}{
        \begin{tikzpicture}[]
            \node (f) [squ, draw] at (0,0) {$f$};
            \node (g) [right = 0.4cm of f] [squ, draw] {$g$};
            \node (dif) [circle, draw, minimum size = 9mm] at (f) {};
            \node (tail) at ($(dif)+(0,-1.4)$) [inner sep = 2pt] {\imarker{$i$}};
            \draw (dif)--(tail);
        \end{tikzpicture}
        }
        &
        +
        &
        \adjustbox{valign=c, scale=1}{
        \begin{tikzpicture}[]
            \node (f) [squ, draw] at (0,0) {$f$};
            \node (g) [right = 0.4cm of f] [squ, draw] {$g$};
            \node (dif) [circle, draw, minimum size = 9mm] at (g) {};
            \node (tail) at ($(dif)+(0,-1.4)$) [inner sep = 2pt] {\imarker{$i$}};
            \draw (dif)--(tail);
        \end{tikzpicture}
        }
        \\
        {}&{}&{\updownarrow}&{}&{}
        \\
        {\partial_i(fg)}
        &
        =
        &
        {\partial_i(f)\,g}
        &
        +
        &
        {f\,\partial_i(g)}
    \end{array}}
\end{align}
This ``differentiation hook'' design was previously suggested by Penrose.\cite{penroserindler1987spinorsandspace-time,penrose2004roadtoreality}
However, he has not published how to do the Euclidean vector calculus in three dimensions using it. As you will see soon, it is powerful to distinguish vector algebraic manipulations from the range of differentiation when an index-free format is kept, while both are denoted without distinction by parentheses in the ordinary notation.

The Leibniz rule, Eq. \ref{eq:Leibnizrule}, can be applied regardless of the operand type.\cite{Note4} For instance, a vector can be fed to $\nabla$.

{\color{white}.}\vspace{-1.5\baselineskip}
\begin{equation}
    \adjustbox{valign=c}{\begin{tikzpicture}
        \node (A) [squ, draw] at (0,0) {$A$};
        \node (dif) [empty] at ($(A)$) {\includegraphics[scale=1.0]{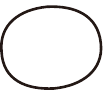}};
        \node (j) [inner sep = 2pt, outer sep = 0] at ($(A)+(0,-1.35)$) {\imarker{$j$}};
        \node (i) [inner sep = 2pt, outer sep = 0] at ($(j)-(0.4,0)$) {\imarker{$i$}};
        \draw (A) to (j);
        \draw (i) -- ($(i)+(0,1)$);
    \end{tikzpicture}}
    \,\,=\,\,
    \partial_i A_j
    \,\,=\,\,
    \big(\text{``}\nabla \vec{A}\text{''}\big)_{ij}
    \label{eq:delA}
\end{equation}
Here, visual reasoning comes earlier, naturally suggesting the concept ``$\nabla\vec{A}$'' without reference to coordinates (before we attach index markers). This is one of the instances where the graphical notation intuitively hints students, who do not have abstract and rigorous mathematical understanding, to enter the world of tensors with its coordinate-free nature unspoiled.

The expression Eq. \ref{eq:delA} can be physically or geometrically meaningful, but it frequently appears in a particular encoding: divergence and curl.\cite{Note5} They are obtained when we let the two tails of Eq. \ref{eq:delA} ``interact'' with each other with the machines we have seen in \cref{sec:Algebra}.
\begin{equation}
    \adjustbox{valign=c}{\begin{tikzpicture}
        \node (shift) [empty] at (0,0.2) {};
        \node (A) [squ, draw] at (0,0) {$A$};
        \node (dif) [empty] at ($(A)+(0.04,0)$) {\includegraphics[scale=1.0]{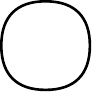}};
        \node (j) [empty] at ($(A)+(0,-0.95)+(shift)$) {};
        \node (i) [empty] at ($(j)+(0.45,0)$) {};
        \draw (A) to (j);
        \node (aux) [outer sep = 0] at ($(i)+(0,0.55)-(shift)$) {};
        \draw ($(dif)-(-0.35,0.3)$) ..controls(aux).. (i);
        \node (mij) [empty] at ($(i)!0.5!(j)$) {};
        \node (mb) [empty] at ($(mij)+(0,-0.225)$) {};
        \draw [rounded corners = 0.225 cm] (i) |- (mb) -| (j);
        \node (phantom) [empty] at ($(0,-0.95)+(0,-0.225)$) {};
    \end{tikzpicture}}
    \,\,=\,\,
    \nabla\cdot\vec{A}\,\,,\quad
    \adjustbox{valign=c}{\begin{tikzpicture}
        \node (shift) [empty] at (0,0.13) {};
        \node (A) [squ, draw] at (0,0) {$A$};
        \node (dif) [empty] at ($(A)-(-0.04,0)$) {\includegraphics[scale=1.0]{figs/CCSqBalloon9mm.pdf}};
        \node (j) [empty] at ($(A)+(0,-0.95)$) {};
        \node (xJ) [empty] at ($(A)+(0,-0.95)+(0,0.1125)$) {};
        \node (i) [empty] at ($(j)-(-0.45,0)$) {};
        \node (xI) [empty] at ($(j)-(-0.45,0)+(0,0.1125)$) {};
        \node (I) [empty] at ($(xI)+(0,0.1125)$) {};
        \node (J) [empty] at ($(xJ)+(0,0.1125)$) {};
        \draw (A) to ($(J)+(shift)$);
        \node (aux) [outer sep = 0] at ($(xI)+(0,0.55)-(0,0.1125)$) {};
        \draw ($(dif)-(-0.35,0.3)$) ..controls(aux).. ($(I)+(shift)$);
        \node (mij) [empty] at ($(i)!0.5!(j)$) {};
        \node (mb) [empty] at ($(mij)+(0,-0.225)$) {};
        \node (mijup) [empty] at ($(mij)+(shift)$) {};
        \draw (mijup)--(mb);
        \draw ($(I)+(shift)$) ..controls($(xI)-({0.01*sqrt(3)},0.01)+(shift)$)..(mijup);
        \draw ($(J)+(shift)$) ..controls($(xJ)-(-{0.01*sqrt(3)},0.01)+(shift)$)..(mijup);
        \fill [fill = black] (mijup) circle (1.3pt);
    \end{tikzpicture}}
    \,\,=\,\,
    \nabla\times\vec{A}\,\,.
    \label{eq:divcurl}
\end{equation}

A final note: the differentiation apply only on boxes, not lines. It is because $\delta_{ij}$'s and $\epsilon_{ijk}$'s are all constants. So, one can freely rearrange the balloons (differentiation) relative to connecting lines and cross product machines regardless of how they are entangled with each other. An imagery that the balloon membrane is impermeable to boxes but do not care whether lines or cross product machines pass through can be helpful.

\subsection{First Derivative Identities}
Finally, we will now show how easy deriving vector calculus identities is with the graphical notation! Essential examples are demonstrated; the remaining identities are worked on the supplementary material\cite{SuppMat} as exercises.
\subsubsection{\texorpdfstring{$\nabla\cdot\big(\vec{A}\times\vec{B}\big)$}{div (A cross B)}}
From the diagrams for the cross product (Eq. \ref{eq:crossproduct}) and the divergence of a vector field (Eq. \ref{eq:divcurl}), $\nabla\cdot\big(\vec{A}\times\vec{B}\big)$ can be easily represented graphically. Then, apply the Leibniz rule Eq. \ref{eq:Leibnizrule}.
\begin{align}
    \adjustbox{valign=c}{
    \begin{tikzpicture}[]
        \node (A) at (-0.3,0) [squ, draw] {$B$};
        \node (B) at (0.3,0) [squ, draw] {$A$};
        \node (cross) at (0,-0.33) [empty] {};
        \fill [fill = black] (cross) circle (1.3pt);
        \node (dif) [empty] at (0,0) {\includegraphics[scale=1.0]{figs/CCSqBalloon9x14mm.pdf}};
        \node (difp) [empty] at (-0.4,-0.4) {};
        \node (end) at (-0.2,-0.8) [empty] {};
        \draw (A) to [out = -60, in = 180] (cross) to [out = 0, in = -120] (B);
        \draw [rounded corners = 0.2 cm] (difp)|- (end)-|(cross);
    \end{tikzpicture}}
    \  &= \
    \adjustbox{valign=c}{\begin{tikzpicture}[]
        \node (A) at (-0.4,0) [squ, draw] {$B$};
        \node (B) at (0.4,0) [squ, draw] {$A$};
        \node (cross) at (0,-0.33) [empty] {};
        \fill [fill = black] (cross) circle (1.3pt);
        \node (dif) at (A) [draw, circle, minimum size = 0.9cm] {};
        \node (end) at (-0.2,-0.8) [empty] {};
        \draw (A) to [out = -45, in = 150] (cross) to [out = 30, in = -135] (B);
        \draw [rounded corners = 0.2 cm] (cross)|-(end)-|(dif);
    \end{tikzpicture}}
    \ + \
    \adjustbox{valign=c}{\begin{tikzpicture}[]
        \node (A) at (-0.4,0) [squ, draw] {$B$};
        \node (B) at (0.4,0) [squ, draw] {$A$};
        \node (cross) at (0,-0.33) [empty] {};
        \fill [fill = black] (cross) circle (1.3pt);
        \node (dif) at (B) [draw, circle, minimum size = 0.9cm] {};
        \draw (A) to [out = -45, in = 150] (cross) to [out = 30, in = -135] (B);
        \node (end) at (0.2,-0.8) [empty] {};
        \draw [rounded corners = 0.2 cm] (cross)|-(end)-|(dif);
    \end{tikzpicture}}
\end{align}
The second term is a contraction of
\adjustbox{valign=c, scale=0.8}{\begin{tikzpicture}[]
    \node (A) at (-0.7,0) [squ, draw] {$B$};
    \node (O) at (0,0) [empty] {};
    \draw (A)--(O);
\end{tikzpicture}}
and
\adjustbox{valign=c, scale=0.8}{\begin{tikzpicture}[]
    \node (B) at (0.4,0) [squ, draw] {$A$};
    \node (cross) at (-0.2,-0.1) [empty] {};
    \node (A) at (-0.6,0) [empty] {};
    \fill [fill = black] (cross) circle (1.3pt);
    \node (dif) at (B) [draw, circle, minimum size = 0.9cm] {};
    \draw (A)--($(A)+(0.03,0)$) to [out=0, in = 160] (cross) to [out=20, in = 180, looseness=0.8] (B);
    \draw (cross) to [out=-90, in=-135, looseness=1.5] (dif);
\end{tikzpicture}}\,, which is $\vec{B}\cdot(\nabla\times\vec{A})$. The first term is a contraction of
$
\adjustbox{valign=c, scale=0.8}{\begin{tikzpicture}[]
    \node (B) at (-0.4,0) [squ, draw] {$B$};
    \node (cross) at (0.2,-0.1) [empty] {};
    \node (A) at (0.6,0) [empty] {};
    \fill [fill = black] (cross) circle (1.3pt);
    \node (dif) at (B) [draw, circle, minimum size = 0.9cm] {};
    \draw (A)--($(A)-(0.03,0)$) to [out=180, in = 20] (cross) to [out=160, in = 0, looseness=0.8] (B);
    \draw (cross) to [out=-90, in=-45, looseness=1.5] (dif);
\end{tikzpicture}}
=-\,\,
\adjustbox{valign=c, scale=0.8}{\begin{tikzpicture}[]
    \node (B) at (-0.4,0) [squ, draw] {$B$};
    \node (cross) at (0.2,-0.1) [empty] {};
    \node (A) at (0.6,0) [empty] {};
    \fill [fill = black] (cross) circle (1.3pt);
    \node (dif) at (B) [draw, circle, minimum size = 0.9cm] {};
    \draw (A)--($(A)-(0.03,0)$) to [out=180, in = 20] (cross) to [out=-180, in = 0, looseness=1.5] (B);
    \draw (cross) to [out=80, in=20, looseness=1.3] (dif);
\end{tikzpicture}}
$
and
\adjustbox{valign=c, scale=0.8}{\begin{tikzpicture}[]
    \node (A) at (0.7,0) [squ, draw] {$A$};
    \node (O) at (0,0) [empty] {};
    \draw (A)--(O);
\end{tikzpicture}}\,,
which is $(-\nabla\times\vec{B})\cdot\vec{A}$. Thus, we obtain $\vec{B}\cdot(\nabla\times\vec{A})-\vec{A}\cdot(\nabla\times\vec{B})$.
We do not need to memorize the tricky minus sign or look up a vector identity list all the time. All we need to do is just to doodle the diagrams and see what happens.

\subsubsection{\texorpdfstring{$\nabla\times\big(\vec{A}\times\vec{B}\big)$}{curl (A cross B)}}
$\nabla\times\big(\vec{A}\times\vec{B}\big)$ can readily be written in a graphical form from the diagrams for the cross product (Eq. \ref{eq:crossproduct}) and a curl of a vector field (Eq. \ref{eq:divcurl}).
The formula is rather complex-looking: $\nabla\times\big(\vec{A}\times\vec{B}\big) = (\nabla\cdot\vec{B})\vec{A} + (\vec{B}\cdot\nabla)\vec{A} - (\nabla\cdot\vec{A})\vec{B} - (\vec{A}\cdot\nabla)\vec{B}$. While proving this in the index notation, you may frown at equations to recognize which indices corresponds to which epsilon and delta; however, it is much neater in the graphical notation.
The proof proceeds by applying the Leibniz rule Eq. \ref{eq:Leibnizrule} and the ``$\,\includegraphics[scale=1.0,valign=c]{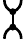}=\includegraphics[scale=1.0,valign=c]{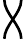}-\includegraphics[scale=1.0,valign=c]{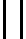}\,$'' identity Eq. \ref{eq:theidentity}.
\begin{align}
    &\adjustbox{valign=c}{\begin{tikzpicture}[]
        \node (A) at (-0.3,0) [squ, draw] {$B$};
        \node (B) at (0.3,0) [squ, draw] {$A$};
        \node (cross1) at (0,-0.3) [empty] {};
        \node (cross2) at (0,-0.6) [empty] {};
        \node (end) at (0, -1) [empty] {};
        \fill [fill = black] (cross1) circle (1.3pt);
        \fill [fill = black] (cross2) circle (1.3pt);
        \node (dif) at (0,0) [empty] {\includegraphics[scale=1.0]{figs/CCSqBalloon9x14mm.pdf}};
        \node (difp) at ($(dif)+(0.33,-0.4)$) [empty] {};
        \draw (A) to [out = -60, in = 180] (cross1) to [out = 0, in = -120] (B);
        \draw (cross1) to (cross2);
        \draw (cross2) to [out = -30, in = -90] (difp);
        \draw (cross2) to (end);
    \end{tikzpicture}}
    =
    \adjustbox{valign=c}{\begin{tikzpicture}[]
        \node (A) at (-0.3,0) [squ, draw] {$B$};
        \node (B) at (0.3,0) [squ, draw] {$A$};
        \node (cross1) at (0,-0.3) [empty] {};
        \node (cross2) at (0,-0.6) [empty] {};
        \node (end) at (0, -1) [empty] {};
        \node (dif) at (0,0) [empty] {\includegraphics[scale=1.0]{figs/CCSqBalloon9x14mm.pdf}};
        \node (difp) at ($(dif)+(0.33,-0.4)$) [empty] {};
        \draw (A) ..controls($(A)+(0,-0.83)$)and($(difp)+(0,-0.4)$).. (difp);
        \draw (B) to [out = -120, in = 90] (end);
    \end{tikzpicture}}
    -
    \adjustbox{valign=c}{\begin{tikzpicture}[]
        \node (A) at (-0.3,0) [squ, draw] {$B$};
        \node (B) at (0.3,0) [squ, draw] {$A$};
        \node (cross1) at (0,-0.3) [empty] {};
        \node (cross2) at (0,-0.6) [empty] {};
        \node (end) at (0, -1) [empty] {};
        \node (dif) at (0,0) [empty] {\includegraphics[scale=1.0]{figs/CCSqBalloon9x14mm.pdf}};
        \node (difp) at ($(dif)+(0.33,-0.4)$) [empty] {};
        \draw (B) ..controls($(B)+(-0.35,-0.85)$)and($(difp)+(0,-0.4)$).. (difp);
        \draw (A) to [out = -90, in = 90] (end);
    \end{tikzpicture}}
    \\
    &=
    \adjustbox{valign=c}{\begin{tikzpicture}[]
        \node (A) at (-0.4,0) [squ, draw] {$B$};
        \node (B) at (0.4,0) [squ, draw] {$A$};
        \node (cross1) at (0,-0.3) [empty] {};
        \node (cross2) at (0,-0.6) [empty] {};
        \node (end) at (0, -1) [empty] {};
        \node (dif) at (A) [draw, circle, minimum size = 0.9cm] {};
        \draw (A) to [out = -90, in = -60, looseness = 4.25] (dif);
        \draw (B) to [out = -90, in = 90] (end);
    \end{tikzpicture}}
    +
    \adjustbox{valign=c}{\begin{tikzpicture}[]
        \node (A) at (-0.4,0) [squ, draw] {$B$};
        \node (B) at (0.4,0) [squ, draw] {$A$};
        \node (cross1) at (0,-0.3) [empty] {};
        \node (cross2) at (0,-0.6) [empty] {};
        \node (end) at (0, -1) [empty] {};
        \node (dif) at (B) [draw, circle, minimum size = 0.9cm] {};
        \draw (A) to (dif);
        \draw (B) to [out = -90, in = 90] (end);
    \end{tikzpicture}}
    -
    \adjustbox{valign=c}{\begin{tikzpicture}[]
        \node (A) at (-0.4,0) [squ, draw] {$B$};
        \node (B) at (0.4,0) [squ, draw] {$A$};
        \node (cross1) at (0,-0.3) [empty] {};
        \node (cross2) at (0,-0.6) [empty] {};
        \node (end) at (0, -1) [empty] {};
        \node (dif) at (B) [draw, circle, minimum size = 0.9cm] {};
        \draw (B) to [out = -90, in = -60, looseness = 4.25] (dif);
        \draw (A) to [out = -90, in = 90] (end);
    \end{tikzpicture}}
    -
    \adjustbox{valign=c}{\begin{tikzpicture}[]
        \node (A) at (-0.4,0) [squ, draw] {$B$};
        \node (B) at (0.4,0) [squ, draw] {$A$};
        \node (cross1) at (0,-0.3) [empty] {};
        \node (cross2) at (0,-0.6) [empty] {};
        \node (end) at (0, -1) [empty] {};
        \node (dif) at (A) [draw, circle, minimum size = 0.9cm] {};
        \draw (B) to (dif);
        \draw (A) to [out = -90, in = 90] (end);
    \end{tikzpicture}}\nonumber
\end{align}
Translating back to the ordinary notation gives the desired result. Note that the second term in the bottom line translates into $(\vec{B}\cdot\nabla)\vec{A}$, since
\adjustbox{valign=c, scale = 0.8}{\begin{tikzpicture}[]
    \node (A) at (-0.45,0) [squ, draw] {$B$};
    \node (B) at (0.4,0) [empty] {$(\cdots)$};
    \node (cross1) at (0,-0.3) [empty] {};
    \node (cross2) at (0,-0.6) [empty] {};
    \node (dif) at (B) [draw, circle, minimum size = 0.9cm] {};
    \draw (A) to (dif);
\end{tikzpicture}}
is the derivative ``modified'' by $\vec{B}$: it ``$\vec{B}$-likely'' differentiates
\raisebox{0.33ex}{\adjustbox{valign=c, scale=0.8}{{$(\cdots)$}}}, that is, the directional derivative with respect to $\vec{B}$, $B_i\partial_i$\raisebox{0.33ex}{\adjustbox{valign=c, scale=0.8}{{$(\cdots)$}}}.

\begin{figure}[t]
    \centering
    \includegraphics[width=1.0\linewidth]{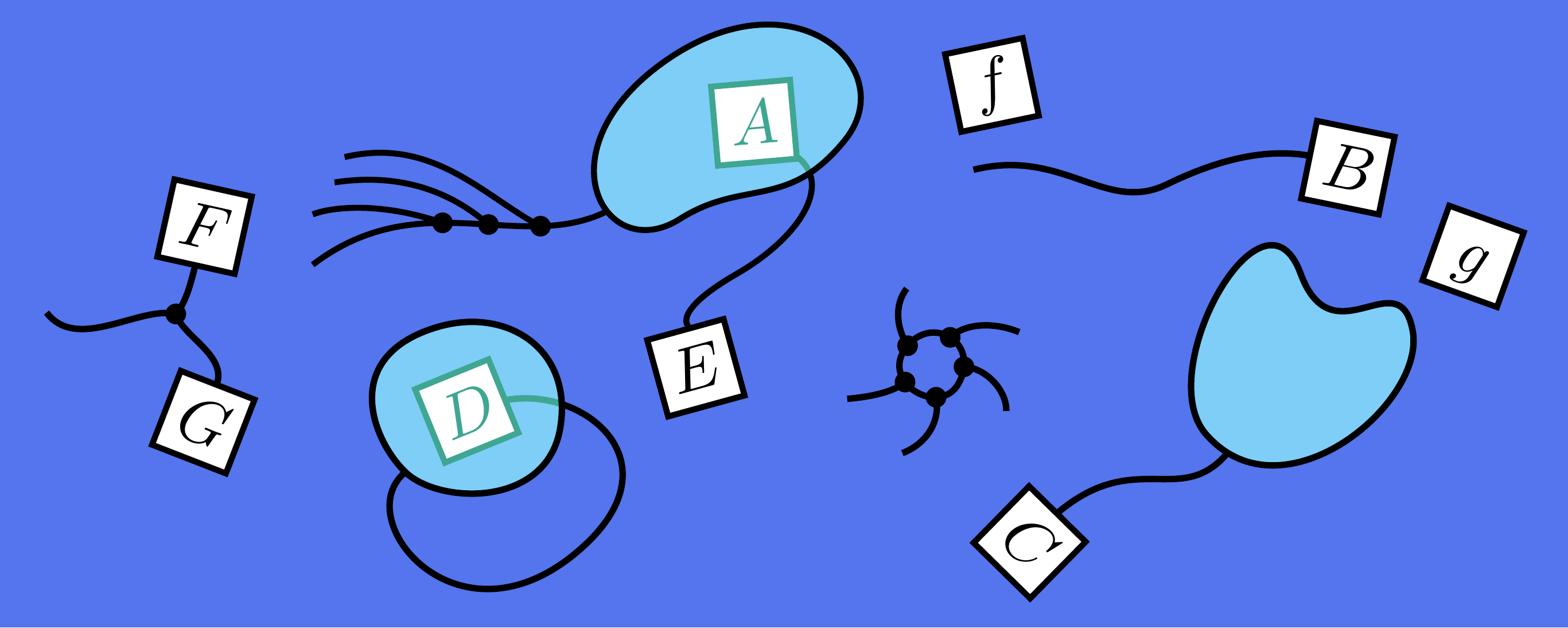}
    \caption{The ``ecosystem'' of the graphical vector calculus.}
    \label{fig:ecosystem}
\end{figure}

\subsubsection{\texorpdfstring{$\nabla\big(\vec{A}\cdot\vec{B}\big)$}{grad (A dot B)}\label{sec:delAdotB}}
Lastly, we will demonstrate a graphical reasoning on the notorious vector calculus identity: $\nabla\big(\vec{A}\cdot\vec{B}\big)$. The formula is given by Eq. \ref{eq:delAdotB}. It is perhaps the most complicated among all vector calculus identities. However, a bigger problem is that it is not clear how to massage $\nabla\big(\vec{A}\cdot\vec{B}\big)$ into smaller expressions. In the graphical notation, one can see the motivation of each step more transparently. Start from the diagram for $\nabla\big(\vec{A}\cdot\vec{B}\big)$:
\begin{equation}
    \label{eq:delAdotBLeibniz}
    \adjustbox{valign=c}{\begin{tikzpicture}[]
        \node (O) at (0,0) [empty] {};
        \node (A) at (0,0.3) [squ,draw] {$A$};
        \node (B) at (0,-0.3) [squ,draw] {$B$};
        \node (dif) at (O) [empty] {\includegraphics[angle=90,origin=c]{figs/CCSqBalloon9x14mm.pdf}};
        \node (difp) at ($(dif)-(0.3,0.58)$) [empty] {};
        \draw (A)--(B);
        \node (I) at (-0.3876,-1) [empty] {};
        \node (i) at (-0.3876,-1.1) [empty] {};
        \draw (difp)..controls($(I)+(0,0.2)$)..(I);
        \draw (I)--(i);
    \end{tikzpicture}}
    \,\,=\,\,
    \adjustbox{valign=c}{\begin{tikzpicture}[]
        \node (O) at (0,0) [empty] {};
        \node (A) at (0,0.3) [squ,draw] {$A$};
        \node (B) at (0,-0.58) [squ,draw] {$B$};
        \node (dif) at (A) [empty] {\includegraphics[scale=1.0, valign=c]{figs/TCCSqBalloon9mmL.pdf}};
        \node (difp) at ($(dif)-(0.3876,0.3692)$) [empty] {};
        \draw (A)--(B);
        \node (i) at (-0.3876,-1.1) [empty] {};
        \draw (difp)--(i);
    \end{tikzpicture}}
    +
    \adjustbox{valign=c}{\begin{tikzpicture}[]
        \node (O) at (0,0) [empty] {};
        \node (A) at (0,0.3) [squ,draw] {$B$};
        \node (B) at (0,-0.58) [squ,draw] {$A$};
        \node (dif) at (A) [empty] {\includegraphics[scale=1.0, valign=c]{figs/TCCSqBalloon9mmL.pdf}};
        \node (difp) at ($(dif)-(0.3876,0.3692)$) [empty] {};
        \draw (A)--(B);
        \node (i) at (-0.3876,-1.1) [empty] {};
        \draw (difp)--(i);
    \end{tikzpicture}}
    \,.
\end{equation}
We aim to express Eq. \ref{eq:delAdotBLeibniz} in tractable terms; we must transform it into vectorial terms that can be written in a coordinate-free manner in the ordinary notation (such as divergence, curl, or directional derivatives).\cite{Note6}
The second term in the right hand side is identical to the first term if $A$ is substituted to $B$ and $B$ is substituted to $A$; therefore, we may work on the first term first then simply do the substitution to obtain the result for the second term.

The central observation that guides us is that if the first term was     \adjustbox{valign=c, scale=0.8}{\begin{tikzpicture}[]
    \node (O) at (0,0) [empty] {};
    \node (A) at (0,0.3) [squ,draw] {$A$};
    \node (B) at (-0.3876,-0.58) [squ,draw] {$B$};
    \node (dif) at (A) [empty] {\includegraphics[scale=1.0, valign=c]{figs/TCCSqBalloon9mmL.pdf}};
    \node (difp) at ($(dif)-(0.3876,0.3692)$) [empty] {};
    \draw (difp)--(B);
    \node (i) at (0,-1.1) [empty] {};
    \draw (A)--(i);
\end{tikzpicture}}, it can be written as $(\vec{B}\cdot\nabla)\vec{A}$. Then, interchanging two lines is readily possible by $\,\includegraphics[valign=c]{figs/Inline_ll.pdf}=\includegraphics[valign=c]{figs/Inline_X.pdf}-\includegraphics[valign=c]{figs/Inline_YY.pdf}\,$.
\begin{equation}
    {\renewcommand{\arraystretch}{1.5}
    \begin{array}{c @{{}={}} c c l}
        \adjustbox{valign=c}{\begin{tikzpicture}[]
            \node (O) at (0,0) [empty] {};
            \node (A) at (0,0.3) [squ,draw] {$A$};
            \node (B) at (0,-0.9) [squ,draw] {$B$};
            \node (dif) at (A) [empty] {\includegraphics[scale=1.0, valign=c]{figs/TCCSqBalloon9mmL.pdf}};
            \node (difp) at ($(dif)-(0.3876,0.3692)$) [empty] {};
            \draw (A)--(B);
            \node (i) at (-0.3876,-1.4) [empty] {};
            \draw (difp)--(i);
        \end{tikzpicture}}
        &
        \hspace{1.1500em}
        \adjustbox{valign=c}{\begin{tikzpicture}[]
            \node (O) at (0,0) [empty] {};
            \node (A) at (0,0.3) [squ,draw] {$A$};
            \node (B) at (0, -0.9) [squ,draw] {$B$};
            \node (dif) at (A) [empty] {\includegraphics[scale=1.0, valign=c]{figs/TCCSqBalloon9mmL.pdf}};
            \node (difp) at ($(dif)-(0.3876,0.3692)$) [empty] {};
            \draw (difp) to[out=-90,in=90] (B);
            \node (i) at (-0.3876,-1.4) [empty] {};
            \node (I) at ($(i)+(0,0.5)$) [empty] {};
            \draw (A) to[out=-90,in=90] (I)--(i);
        \end{tikzpicture}}
        \hspace{-1.1500em}
        &-
        &
        \hspace{0.3em}\hspace{1.11em}
        \adjustbox{valign=c}{\begin{tikzpicture}[]
            \node (O) at (0,0) [empty] {};
            \node (A) at (0,0.3) [squ,draw] {$A$};
            \node (B) at (-0.3876,-0.9) [empty] {};
            \node (b) at (0,-0.9) [squ,draw] {$B$};
            \node (M) at ($(A)!0.5!(B)$) [empty] {};
            \node (c1) at ($(M)+(0,0.0)$) [empty] {};
            \node (c2) at ($(M)-(0,0.2)$) [empty] {};
            \fill [fill = black] (c1) circle (1.3pt);
            \fill [fill = black] (c2) circle (1.3pt);
            \node (dif) at (A) [empty] {\includegraphics[scale=1.0, valign=c]{figs/TCCSqBalloon9mmL.pdf}};
            \node (difp) at ($(dif)-(0.3876,0.3692)$) [empty] {};
            \draw (difp) to [out=-90, in=150] (c1) to [out=30, in=-90] ($(difp)+(0.3876,0)$) --(A);
            \draw (c1)--(c2);
            \node (i) at (-0.3876,-1.4) [empty] {};
            \draw (b) to [out=90, in=-30] (c2) to [out=210, in=90] (B) --(i);
        \end{tikzpicture}}
        \\
        \phantom{.}\hspace{3.3em}\phantom{.}
        &
        \hspace{1.1500em}
        \adjustbox{valign=c}{\begin{tikzpicture}[]
            \node (O) at (0,0) [empty] {};
            \node (A) at (0,0.3) [squ,draw] {$A$};
            \node (B) at (-0.3876,-0.9) [squ,draw] {$B$};
            \node (dif) at (A) [empty] {\includegraphics[scale=1.0, valign=c]{figs/TCCSqBalloon9mmL.pdf}};
            \node (difp) at ($(dif)-(0.3876,0.3692)$) [empty] {};
            \draw (difp)--(B);
            \node (i) at (0,-1.4) [empty] {};
            \draw (A)--(i);
        \end{tikzpicture}}
        \hspace{-1.1500em}
        &+
        &
        \hspace{0.3em}
        \adjustbox{valign=c}{\begin{tikzpicture}[]
            \node (O) at (0,0) [empty] {};
            \node (a) at (-0.3876,0.3) [squ,draw] {$A$};
            \node (A) at (0,0.3) [empty] {};
            \node (B) at (-0.3876,-0.9) [empty] {};
            \node (b) at (0,-0.9) [squ,draw] {$B$};
            \node (M) at ($(A)!0.5!(B)$) [empty] {};
            \node (c1) at ($(M)+(0,0.0)$) [empty] {};
            \node (c2) at ($(M)-(0,0.2)$) [empty] {};
            \fill [fill = black] (c1) circle (1.3pt);
            \fill [fill = black] (c2) circle (1.3pt);
            \node (dif) at (a) [empty] {\includegraphics[scale=1.0, valign=c]{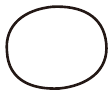}};
            \node (difp) at ($(dif)-(-0.3876,0.315)$) [empty] {};
            \draw (difp) to [out=-90, in=30] (c1) to [out=150, in=-90] ($(difp)+(-0.3876,0)$) --(a);
            \draw (c1)--(c2);
            \node (i) at (-0.3876,-1.4) [empty] {};
            \draw (b) to [out=90, in=-30] (c2) to [out=210, in=90] (B) --(i);
        \end{tikzpicture}}
        \\
        &\hspace{0.42em}(\vec{B}\cdot\nabla)\vec{A}\,
        &+
        &\vec{B}\times(\nabla\times\vec{A})
    \end{array}}
\end{equation}
In the second line, the upper cross product machine is ``clanked.'' Finally,
\begin{equation}
    \label{eq:delAdotB}
    \nabla\big(\vec{A}\cdot\vec{B}\big)
    =
    (\vec{B}\cdot\nabla) \vec{A}
    + \vec{B} \times (\nabla\times\vec{A})
    + (\vec{A}\leftrightarrow \vec{B}),
\end{equation}
where ``$+(\vec{A}\leftrightarrow \vec{B})$'' means adding the same expression with $\vec{A}$ and $\vec{B}$ interchanged.
This trick of interchanging two lines, $\,\includegraphics[valign=c]{figs/Inline_ll.pdf}=\includegraphics[valign=c]{figs/Inline_X.pdf}-\includegraphics[valign=c]{figs/Inline_YY.pdf}\,$, is often useful. With the graphical notation, utilizing it and recognizing when to use it is achieved without difficulty.

\subsection{Second and Higher-order Derivative Identities}
Graphical proofs of second and higher order identities can be easily proceeded analogously.
Second-order derivatives are depicted as double-balloon diagrams. There are no new graphical rules introduced except the following ``commutativity of derivatives,''
\begin{equation}
    \label{eq:commutativityofderivatives}
    \adjustbox{valign=c}{
    \begin{tikzpicture}[]
    	\node (f) at (0,0) [] {};
    	\draw (0.2,0)--(0.2,-0.9);
        \node (dif2) at (f) [circle, minimum size = 9mm, draw,fill=white] {};
    	\draw (-0.2,0)--(-0.2,-0.9);
        \node (dif1) at (f) [circle, minimum size = 6.5mm, draw,fill=white] {};
    \end{tikzpicture}}
    \ = \
    \adjustbox{valign=c}{
    \begin{tikzpicture}[]
    	\node (f) at (0,0) [] {};
    	\draw (-0.2,0)--(-0.2,-0.9);
        \node (dif2) at (f) [circle, minimum size = 9mm, draw,fill=white] {};
    	\draw (0.2,0)--(0.2,-0.9);
        \node (dif1) at (f) [circle, minimum size = 6.5mm, draw,fill=white] {};
    \end{tikzpicture}}
    \,\,,
\end{equation}
where anything smooth that the derivatives commute can be go inside the balloons. This is translated into the ordinary notation as $\partial_j\partial_i=\partial_i\partial_j$ as an operator identity.

One of the most immediate results in second order derivatives is the following.
\begin{equation}
    \adjustbox{valign=c}{
    \begin{tikzpicture}[]
    	\node (f) at (0,0) [] {};
        \node (dif1) at (f) [circle, minimum size = 6.5mm, draw] {};
        \node (dif2) at (f) [circle, minimum size = 9mm, draw] {};
        \node (cross) at (0,-0.75) [empty] {};
        \fill (cross) circle (1.3pt);
        \node (end) at (0, -1.1) [empty] {};
        \draw (cross) to [out = 160, in = -120] (dif1);
        \draw (cross) to [out = 20, in = -68] (dif2);
        \draw (cross) to (end);
    \end{tikzpicture}}
    \ = \
    \adjustbox{valign=c}{
    \begin{tikzpicture}[]
    	\node (f) at (0,0) [] {};
        \node (dif1) at (f) [circle, minimum size = 6.5mm, draw] {};
        \node (dif2) at (f) [circle, minimum size = 9mm, draw] {};
        \node (cross) at (0,-0.75) [empty] {};
        \fill (cross) circle (1.3pt);
        \node (end) at (0, -1.1) [empty] {};
        \draw (cross) to [out = 160, in = -112] (dif2);
        \draw (cross) to [out = 20, in = -60] (dif1);
        \draw (cross) to (end);
    \end{tikzpicture}}
    \ = \
    \adjustbox{valign=c}{
    \begin{tikzpicture}[]
    	\node (f) at (0,0) [] {};
        \node (dif1) at (f) [circle, minimum size = 6.5mm, draw] {};
        \node (dif2) at (f) [circle, minimum size = 9mm, draw] {};
        \node (cross) at (0,-0.75) [empty] {};
        \fill (cross) circle (1.3pt);
        \node (end) at (0, -1.1) [empty] {};
        \node (int) at ($(cross)+(0,0.17)$) [empty] {};
        \draw (cross) to[out=160, in=210,looseness=2] (int) to[out=30, in=-70, looseness=0.3] (dif2);
        \draw (cross) to[out=20, in=-30,looseness=2] (int) to[out=-210, in=-120, looseness=1] (dif1);
        \draw (cross) to (end);
    \end{tikzpicture}}
    \ = \
    -\adjustbox{valign=c}{
    \begin{tikzpicture}[]
    	\node (f) at (0,0) [] {};
        \node (dif1) at (f) [circle, minimum size = 6.5mm, draw] {};
        \node (dif2) at (f) [circle, minimum size = 9mm, draw] {};
        \node (cross) at (0,-0.75) [empty] {};
        \fill (cross) circle (1.3pt);
        \node (end) at (0, -1.1) [empty] {};
        \draw (cross) to [out = 160, in = -120] (dif1);
        \draw (cross) to [out = 20, in = -68] (dif2);
        \draw (cross) to (end);
    \end{tikzpicture}}
    \ = \
    0
\end{equation}
At the first equality, the inner balloon is rearranged to be the outer one according to Eq. \ref{eq:commutativityofderivatives}; the second equality comes from the \textit{dancing rule}; at the third equality, the ``clank'' process is used.
One can easily see that $\nabla\times(\nabla f)=0$ and $\nabla\cdot(\nabla\times\vec{A})=0$ are all the consequences of this property. The details are contained in the supplementary material\cite{SuppMat} with the proof of other second and higher order identities.

\section{Practical Examples}
So far, this is the story of the graphical notation, a beginners' companion to vector calculus. In this section, we provide practical examples in the physics context.

\subsection{The Economy of the Graphical Notation: The Same Diagram, Different Readings}
Remember \textit{the economy of the graphical notation} in \cref{sec:meetthecrosspoduct}? In music, there are musical objects that have multiple names in ordinary notation. For example, $\mathrm{D}\sharp$ and $\mathrm{E}\flat$ are the same when they are aurally represented.
Likewise, there are situations that different plaintext equations are represented as a single graphical expression so that one can easily recognize their equivalence. The following two, which appears when one deals with the equations of motion of a rigid electric dipole translating and rotating in a magnetic field,\cite{Note7} are equal in their values but spelled differently in the ordinary notation.
\begin{equation}
    \label{eq:aphodipole}
    \vec{v}\cdot\big((\vec{\omega}\times\vec{p})\times\vec{B}\big)
    \,,\ \,
    -\big(\vec{p}\times(\vec{v}\times\vec{B})\big)\cdot\vec{\omega}
\end{equation}
To see the equivalence of them, one should spend time on permuting the vectors according to properties of the triple products. However, it is strikingly easy if one draws a diagram corresponding to them.
\begin{equation}
    \label{eq:aphodipolegrpahical}
    \adjustbox{valign=c}{\begin{tikzpicture}
        \node (O) at (0,0) [empty] {};
        \node (O1) at (-0.3,0) [empty] {};
        \node (O2) at (+0.3,0) [empty] {};
        \node (A) at ($(O1)+(120:0.6)$) [squ,draw] {$\omega$};
        \node (B) at ($(O1)+(240:0.6)$) [squ,draw] {$p$};
        \node (C) at ($(O2)+(-60:0.6)$) [squ,draw] {$B$};
        \node (D) at ($(O2)+(+60:0.6)$) [squ,draw] {$v$};
        \draw (A)--(O1)--(B);
        \draw (C)--(O2)--(D);
        \draw (O1)--(O2);
        \fill (O1) circle (1.3pt);
        \fill (O2) circle (1.3pt);
    \end{tikzpicture}}
\end{equation}
Two expressions in Eq. \ref{eq:aphodipole} are just different readings (groupings) of Eq. \ref{eq:aphodipolegrpahical}. It is the matter of grouping the left branch ($\vec{\omega}\times\vec{p}$) first or the right branch ($\vec{v}\times\vec{B}$) first in Eq. \ref{eq:aphodipolegrpahical}. Permuting the vectors in the ordinary notation and in the graphical notation are just two different ways of manipulating an identical tensor structure, but it is much easier in the graphical notation. Then, why not use the graphical notation, at least as a mnemonic?

\subsection{Cross Your Fingers}
The capacity of the graphical notations is more than a mnemonic. It is a calculation tool equipped with its own syntax so that one can proceed the entire process of vector calculus in the graphical notation without reference to indices. Let us demonstrate such calculational advantages.

The trick of interchanging lines introduced in \cref{sec:delAdotB} has an objective to reassign contractions between indices to obtain a more convenient form. For an example of its practical usage, consider the electrostatic force formula for a point electric dipole $\vec{p}$ in an electric field $\vec{E}(\vec{r})$. It is given by $\nabla\big({\vec{p}\cdot\vec{E}(\vec{r})}\big)$, but also $(\vec{p}\cdot\nabla)\vec{E}(\vec{r})$. It would be an overkill to look up the vector calculus identity table and apply the general formula Eq. \ref{eq:delAdotB}, because $\vec{p}$ is not differentiated by $\nabla$. Simply, the following graphical equations completes the proof of the equivalence of the two.
\begin{equation}
    \adjustbox{valign=c}{\begin{tikzpicture}[]
        \node (O) at (0,0) [empty] {};
        \node (A) at (0,-0.3) [squ,draw] {$E$};
        \node (B) at (0,0.3) [squ,draw] {$p$};
        \node (dif) at (O) [empty] {\includegraphics[angle=90,origin=c]{figs/CCSqBalloon9x14mm.pdf}};
        \node (difp) at ($(dif)-(0,-0.7)$) [empty] {};
        \draw (A)--(B);
        \draw (difp)--(0,1.4);
    \end{tikzpicture}}
    \,=\,
     \adjustbox{valign=c}{\begin{tikzpicture}[]
        \node (O) at (0,0) [empty] {};
        \node (A) at (0,-0.3) [squ,draw] {$E$};
        \node (B) at (0,0.9) [squ,draw] {$p$};
        \node (dif) at (A) [empty] {\includegraphics[scale=1.0, valign=c]{figs/TCCSqBalloon9mmL.pdf}};
        \node (difp) at ($(dif)-(0.3876,-0.2749)$) [empty] {};
        \draw (A)--(B);
        \node (i) at (-0.3876,1.4) [empty] {};
        \draw (difp)--(i);
    \end{tikzpicture}}
    \,=\,
    \adjustbox{valign=c}{\begin{tikzpicture}[]
        \node (O) at (0,0) [empty] {};
        \node (A) at (0,-0.3) [squ,draw] {$E$};
        \node (B) at (-0.3876,0.9) [empty] {};
        \node (p) at (0,0.9) [squ,draw] {$p$};
        \node (dif) at (A) [empty] {\includegraphics[scale=1.0, valign=c]{figs/TCCSqBalloon9mmL.pdf}};
        \node (difp) at ($(dif)-(0.3876,-0.2749)$) [empty] {};
        \node (i) at (-0.3876,1.4) [empty] {};
        \draw (difp) to[out=90,in=-90] (p);
        \draw (A) to[out=90,in=-90] (B)--(i);
    \end{tikzpicture}}\,
    \hspace{-0.2em}-\,
    \adjustbox{valign=c}{\begin{tikzpicture}[]
        \node (O) at (0,0) [empty] {};
        \node (A) at (0,-0.3) [squ,draw] {$E$};
        \node (B) at (-0.3876,0.9) [empty] {};
        \node (p) at (0, 0.9) [squ, draw] {$p$};
        \node (M) at ($(A)!0.5!(B)$) [empty] {};
        \node (c1) at ($(M)+(0,0.0)$) [empty] {};
        \node (c2) at ($(M)-(0,-0.2)$) [empty] {};
        \fill [fill = black] (c1) circle (1.3pt);
        \fill [fill = black] (c2) circle (1.3pt);
        \node (dif) at (A) [empty] {\includegraphics[scale=1.0, valign=c]{figs/TCCSqBalloon9mmL.pdf}};
        \node (difp) at ($(dif)-(0.3876,-0.2749)$) [empty] {};
        \draw (difp) to [out=90, in=-150] (c1) to [out=-30, in=90] ($(difp)+(0.3876,0)$) --(A);
        \draw (c1)--(c2);
        \node (i) at (-0.3876,1.4) [empty] {};
        \draw (i)--(B) to [out=-90, in=-210] (c2) to [out=30, in=-90] (p);
        \draw [thick, red, ->] (-0.5,0.33-0.08)--(0.5,-0.67-0.08);
    \end{tikzpicture}}
\end{equation}
Note that
$\adjustbox{valign=c,scale=0.8}{\begin{tikzpicture}[]
    \node (O) at (0,0) [empty] {};
    \node (A) at (0,-0.3) [squ,draw] {$E$};
    \node (M) at ($(A)!0.5!(B)$) [empty] {};
    \node (c1) at ($(M)+(0,0.0)$) [empty] {};
    \node (c2) at ($(M)-(0,-0.4)$) [empty] {};
    \fill [fill = black] (c1) circle (1.3pt);
    \node (dif) at (A) [empty] {\includegraphics[scale=1.0, valign=c]{figs/TCCSqBalloon9mmL.pdf}};
    \node (difp) at ($(dif)-(0.3876,-0.2749)$) [empty] {};
    \draw (difp) to [out=90, in=-150] (c1) to [out=-30, in=90] ($(difp)+(0.3876,0)$) --(A);
    \draw (c1)--(c2);
\end{tikzpicture}}=\nabla\times\vec{E}(\vec{r})=0$. This shows the intention of the calculation evidently, without memorizing the whole formula. In case of a point magnetic dipole $\vec{m}$ in a magnetic field $\vec{B}(\vec{r})$,
\begin{equation}
    \adjustbox{valign=c}{\begin{tikzpicture}[]
        \node (O) at (0,0) [empty] {};
        \node (A) at (0,-0.3) [squ,draw] {$B$};
        \node (B) at (0,0.3) [squ,draw] {$m$};
        \node (dif) at (O) [empty] {\includegraphics[angle=90,origin=c]{figs/CCSqBalloon9x14mm.pdf}};
        \node (difp) at ($(dif)-(0,-0.7)$) [empty] {};
        \draw (A)--(B);
        \draw (difp)--(0,1.4);
    \end{tikzpicture}}
    \,\,=\,\,
    \adjustbox{valign=c}{\begin{tikzpicture}[]
        \node (O) at (0,0) [empty] {};
        \node (A) at (0,-0.3) [squ,draw] {$B$};
        \node (B) at (-0.3876,0.9) [empty] {};
        \node (m) at (0,0.9) [squ,draw] {$m$};
        \node (dif) at (A) [empty] {\includegraphics[scale=1.0, valign=c]{figs/TCCSqBalloon9mmL.pdf}};
        \node (difp) at ($(dif)-(0.3876,-0.2749)$) [empty] {};
        \node (i) at (-0.3876,1.4) [empty] {};
        \draw (difp) to[out=90,in=-90] (m);
        \draw (A) to[out=90,in=-90] (B)--(i);
    \end{tikzpicture}}
    \hspace{-0.15em}-\,
    \adjustbox{valign=c}{\begin{tikzpicture}[]
        \node (O) at (0,0) [empty] {};
        \node (A) at (0,-0.3) [squ,draw] {$B$};
        \node (B) at (-0.3876,0.9) [empty] {};
        \node (p) at (0, 0.9) [squ, draw] {$m$};
        \node (M) at ($(A)!0.5!(B)$) [empty] {};
        \node (c1) at ($(M)+(0,0.0)$) [empty] {};
        \node (c2) at ($(M)-(0,-0.2)$) [empty] {};
        \fill [fill = black] (c1) circle (1.3pt);
        \fill [fill = black] (c2) circle (1.3pt);
        \node (dif) at (A) [empty] {\includegraphics[scale=1.0, valign=c]{figs/TCCSqBalloon9mmL.pdf}};
        \node (difp) at ($(dif)-(0.3876,-0.2749)$) [empty] {};
        \draw (difp) to [out=90, in=-150] (c1) to [out=-30, in=90] ($(difp)+(0.3876,0)$) --(A);
        \draw (c1)--(c2);
        \node (i) at (-0.3876,1.4) [empty] {};
        \draw (i)--(B) to [out=-90, in=-210] (c2) to [out=30, in=-90] (p);
    \end{tikzpicture}}
    \hspace{0.12em}=\,\,
    \adjustbox{valign=c}{\begin{tikzpicture}[]
        \node (O) at (0,0) [empty] {};
        \node (A) at (0,-0.3) [squ,draw] {$B$};
        \node (B) at (-0.3876,0.9) [squ,draw] {$m$};
        \node (dif) at (A) [empty] {\includegraphics[scale=1.0, valign=c]{figs/TCCSqBalloon9mmL.pdf}};
        \node (difp) at ($(dif)-(0.3876,-0.2749)$) [empty] {};
        \draw (difp)--(B);
        \node (i) at (0,1.4) [empty] {};
        \draw (A)--(i);
    \end{tikzpicture}}
    \hspace{-0.35em}+\,\,\,
    \adjustbox{valign=c}{\begin{tikzpicture}[]
        \node (O) at (0,0) [squ,draw] {\adjustbox{scale=0.9,valign=c}{$\mu_0$}};
    \end{tikzpicture}}
    \adjustbox{valign=c}{\begin{tikzpicture}[]
        \node (O) at (0,0) [empty] {};
        \node (A) at (0,-0.3) [empty] {};
        \node (B) at (-0.3876,0.9) [squ,draw] {$m$};
        \node (M) at ($(A)!0.5!(B)$) [empty] {};
        \node (J) at ($(M)-(0,0.25)$) [squ,draw] {$J$};
        \node (c2) at ($(M)-(0,-0.2)$) [empty] {};
        \fill [fill = black] (c2) circle (1.3pt);
        \draw (J)--(c2);
        \node (i) at (0,1.4) [empty] {};
        \draw (B) to [out=-90, in=-210] (c2) to [out=30, in=-90] ($(B)+(0.3876,0)$) --(i);
        \node (phantom) at (0,-0.75) [empty] {};
    \end{tikzpicture}}\hspace{0.5em},
\end{equation}
so the force exerted on the dipole is $\nabla\big(\vec{m}\cdot\vec{B}(\vec{r})\big) = (\vec{m}\cdot\nabla)\vec{B}(\vec{r})+\mu_0\vec{m}\times\vec{J}(\vec{r})$, where $\vec{J}(\vec{r})=\frac{1}{\mu_0}\nabla\times\vec{B}(\vec{r})$ is current density at $\vec{r}$.
%The trick of interchanging lines has other usage, such as derivation and calculation of magnetic multipole moments.

\subsection{\texorpdfstring{Identities Involving $\vec{r}$}{Identities Involving r}\label{identitiesinvolvingr}}
As a specific and important example, consider the vector calculus with the position vector, $\vec{r}$. First, note that
\begin{equation}
    \adjustbox{valign=c}{\begin{tikzpicture}[]
        \node (r) at (0,0) [squ,draw] {$r$};
        \node (dif) at (r) [circle,draw,minimum size=9mm] {};
        \node (left) at ($(r)+(1,0)$) [empty] {};
        \node (right) at ($(r)+(-1,0)$) [empty] {};
        \draw (left) to (dif);
        \draw (right) to (r);
    \end{tikzpicture}}
    \,=\,
    \adjustbox{valign=c}{\begin{tikzpicture}[]
        \node (r) at (0,0) [] {$\phantom{r}$};
        \node (left) at ($(r)+(-1,0)$) [empty] {};
        \node (right) at ($(r)+(1,0)$) [empty] {};
        \draw (left) to (right);
    \end{tikzpicture}}\,\,,
\end{equation}
which is $\partial_i{x_j}=\delta_{ij}$
in the ordinary notation. If the two terminals are connected by Kronecker delta, a ``vacuum bubble'' is obtained:
\begin{equation}
    \label{eq:vacbubble}
    \adjustbox{valign=c}{\begin{tikzpicture}[]
        \node (phantom) at (0,0.56) [empty] {};
        \node (phantom2) at (0,-1.15) [empty] {};
        \node (O) at (0,0) [empty] {};
        \node (b) at (0,-0.6) [empty] {};
        \node (controlx) at (-0.3,0) [empty] {};
        \node (br) at ($(b)+(-0.2192,0)$) [empty] {};
        \node (bl) at ($(b)-(0.2192,0)$) [empty] {};
        \node (cr) at ($(O)!0.5!(b)+(-0.2192,0)+(controlx)$) [empty] {};
        \node (cl) at ($(O)!0.5!(b)-(-0.2192,0)-(controlx)$) [empty] {};
        \draw [rounded corners = 0.3cm] (O)-|(cl)|-(bl);
        \node (dif) at (O) [circle, draw,fill=white,minimum size=9mm] {};
        \draw [rounded corners = 0.3cm] (O)-|(cr)|-(br);
        \draw (bl)--(br);
        \node (r) at (O) [squ,draw,fill=white] {$r$};
    \end{tikzpicture}}
    \ =\
    \adjustbox{valign=c}{\begin{tikzpicture}[]
        \node (phantom) at (0,0.56) [empty] {};
        \node (phantom2) at (0,-1.15) [empty] {};
        \node (O) at (0,0) [empty] {};
        \node (b) at (0,-0.6) [empty] {};
        \node (controlx) at (0.3,0) [empty] {};
        \node (br) at ($(b)+(0.2192,0)$) [empty] {};
        \node (bl) at ($(b)-(-0.2192,0)$) [empty] {};
        \node (cr) at ($(O)!0.5!(b)+(0.2192,0)+(controlx)$) [empty] {};
        \node (cl) at ($(O)!0.5!(b)-(0.2192,0)-(controlx)$) [empty] {};
        \draw [rounded corners = 0.3cm] (O)-|(cl)|-(bl);
        \draw [rounded corners = 0.3cm] (O)-|(cr)|-(br);
        \draw (bl)--(br);
        \node (i) at (O) [empty,inner sep=1pt] {\raisebox{0.3cm}{\imarker{$i$}}};
        \node (j) at (b) [empty,inner sep=1pt] {\raisebox{-0.55cm}{\imarker{$j$}}};
    \end{tikzpicture}}\,
    \ =\
    \delta_{ij}\delta_{ij}
    \ =\
    \adjustbox{valign=c}{\begin{tikzpicture}
        \node (O) at (0,0) [squ,draw] {$3$};
    \end{tikzpicture}}
    \,\,.
\end{equation}
If a cross product machine is used, \begin{equation}
\adjustbox{valign=c}{\begin{tikzpicture}[]
    \node (r) at (0,0.50) [empty] {};
    \node (c) at (0,-0.15) [empty] {};
    \node (i) at (0,-0.55) [empty] {};
    \fill [fill = black] (c) circle (1.3pt);
    \draw (c) to[out=-75+90,in=-90+90,looseness=2.1] (r);
    \node (dif) at (r) [draw,circle,minimum size=9mm,fill=white] {};
    \draw (c) to[out=75+90,in=90+90,looseness=2.1] (r);
    \node (R) at (r) [squ,draw,fill=white] {$r$};
    \draw (c)--(i);
\end{tikzpicture}}
=
\adjustbox{valign=c}{\begin{tikzpicture}[]
    \node (r) at (0,0.50) [empty] {};
    \node (phantom) at ($(r)+(0,0.45)$) [empty] {};
    \node (c) at (0,-0.15) [empty] {};
    \node (i) at (0,-0.55) [empty] {};
    \fill [fill = black] (c) circle (1.3pt);
    \draw (c) to[out=-75+90,in=-90+90,looseness=2] (r);
    \draw (c) to[out=75+90,in=90+90,looseness=2] (r);
    \draw (c)--(i);
\end{tikzpicture}}
=
\,-
\hspace{-0.55em}
\adjustbox{valign=c}{\begin{tikzpicture}[]
    \node (r) at (0,0.50) [empty] {};
    \node (phantom) at ($(r)+(0,0.45)$) [empty] {};
    \node (c) at (0,-0.15) [empty] {};
    \node (i) at (0,-0.55) [empty] {};
    \fill [fill = black] (c) circle (1.3pt);
    \draw (c) to[out=-75+90,in=+90+90,looseness=2.4] (r);
    \draw (c) to[out=75+90,in=-90+90,looseness=2.4] (r);
    \draw (c)--(i);
\end{tikzpicture}}
\hspace{-0.8em}
\,=
\,-
\adjustbox{valign=c}{\begin{tikzpicture}[]
    \node (r) at (0,0.50) [empty] {};
    \node (phantom) at ($(r)+(0,0.45)$) [empty] {};
    \node (c) at (0,-0.15) [empty] {};
    \node (i) at (0,-0.55) [empty] {};
    \fill [fill = black] (c) circle (1.3pt);
    \draw (c) to[out=-75+90,in=-90+90,looseness=2] (r);
    \draw (c) to[out=75+90,in=90+90,looseness=2] (r);
    \draw (c)--(i);
\end{tikzpicture}}
=0\,\,,
\end{equation}
as you know that $\nabla\times\vec{r}=0$.
The second and the third equality proceed by ``swap-then-yanking'' the cross product machine and the Kronecker delta part, respectively. Lastly, note that
$
    \adjustbox{valign=c,scale=0.8}{\begin{tikzpicture}[]
        \node (r) at (0,0) [squ,draw] {$r$};
        \node (dif) at (r) [circle,draw,minimum size=9mm] {};
        \node (left) at ($(r)+(-1,0)$) [empty] {};
        \draw (left) to (dif);
    \end{tikzpicture}}
    \,=\,
    \adjustbox{valign=c,scale=0.8}{\begin{tikzpicture}[]
        \node (r) at (0,0) [squ,draw] {$n$};
        \node (left) at ($(r)+(-1,0)$) [empty] {};
        \draw (left) to (r);
    \end{tikzpicture}}\,\,,
$
where $\vec{n}:=\vec{r}/r$ ($r:=\mathopen|\vec{r}\hspace{0.1em}\mathclose|$) is the unit radial vector.
%\footnote{In principle, this equation can be derived from $\partial_i x_j = \delta_{ij}$, regarding the left hand side as $\nabla(\sqrt{r^2})$ then applying the chain rule, so it is not a new axiom...}

With these basic graphical equations, one can graphically prove identities involving $r$ and $\vec{r}$ such as the following.
\begin{gather}
    (\vec{A}\hspace{0.15em}\nabla)\,\vec{r}=\vec{A}
    \,\,\,\,\,\leftrightarrow\,\,\,\,\,
    \adjustbox{valign=c}{
    \begin{tikzpicture}[]
        \node (r) at (0,0) [squ,draw] {$r$};
        \node (dif) at (r) [circle,draw,minimum size=9mm] {};
        \node (left) at ($(r)+(-0.9,0)$) [squ,draw] {$A$};
        \node (right) at ($(r)+(0.9,0)$) [empty] {};
        \draw (left) to (dif);
        \draw (right) to (r);
    \end{tikzpicture}}
    \,=\,
    \adjustbox{valign=c}{
    \begin{tikzpicture}[]
        \node (r) at (0,0) [] {$\phantom{r}$};
        \node (left) at ($(r)+(-0.5,0)$) [squ,draw] {$A$};
        \node (right) at ($(r)+(0.5,0)$) [empty] {};
        \draw (left) to (right);
    \end{tikzpicture}}
    \\
    \label{eq:laplacianr}
    \nabla^2\vec{r}=0
    \,\,\,\,\,\leftrightarrow\,\,\,\,\,
    \adjustbox{valign=c}{\begin{tikzpicture}[]
        \node (d1) at (0,0.3) [empty] {};
        \node (d2) at ($(0,0)-(d1)$) [empty] {};
        \node (m) at (-0.8,0) [empty] {};
        \draw [rounded corners = 0.3 cm] (d1) -| (m);
        \node (dif2) at (r) [circle,draw,minimum size=11mm,fill=white] {};
        \draw [rounded corners = 0.3 cm] (m) |- (d2);
        \node (dif) at (r) [circle,draw,minimum size=8.5mm,fill=white] {};
        \node (r) at (0,0) [squ,draw] {$r$};
        \node (right) at ($(r)+(0.9,0)$) [empty] {};
        \draw (right) to (r);
    \end{tikzpicture}}
    \,=\,
    \adjustbox{valign=c}{\begin{tikzpicture}[]
        \node (d1) at (0,0.3) [empty] {};
        \node (m) at (-0.8,0) [empty] {};
        \draw [rounded corners = 0.3 cm] (d1) -| (m);
        \node (dif2) at (r) [circle,draw,minimum size=11mm,fill=white] {};
        \draw (m) to[out=-90,in=180,looseness=1] (a);
        \node (a) at (0.5,0) [empty] {};
        \node (right) at (0.9,0) [empty] {};
        \draw (right) to (a);
    \end{tikzpicture}}
    \,=\,
    0
\end{gather}
Here, the fact that
$
\partial_k{\delta_{ij}}=0\,\leftrightarrow\,
\adjustbox{valign=c,scale=0.8}{\begin{tikzpicture}[]
    \node (i) at (-0.8,-0.1) [empty, inner sep=1pt] {\imarker{$i$}};
    \node (j) at (0.75, 0.05) [empty, inner sep=1pt] {\imarker{$j$}};
    \node (O) at (0,0) [circle,draw,minimum size=5mm] {};
    \node (k) at (-0.66,0.22) [empty, inner sep=1pt] {\imarker{$k$}};
    \draw (i) to[out=0,in=180,looseness=1.2] (j);
    \draw (k) to[out=-20,in=150] (O);
\end{tikzpicture}}
= 0
$
is used.\cite{Note8}
Also, expressions such as $\nabla\times(r\sin\theta\hspace{0.1em}\basise_{\hat{\phi}})$ ($\basise_{\hat{\phi}}:=\nabla\phi/\mathopen|\nabla\phi\mathclose|$, where $\phi$ is the azimuthal angle) can be calculated by recasting it into a coordinate-free expression: $\nabla\times(\basise_{z}\times\vec{r})$.
\begin{equation}
    \includegraphics[scale=1.0,valign=c]{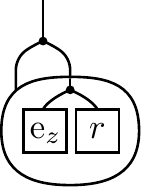}
    \hspace{0.06em}=\hspace{0.06em}
    \includegraphics[scale=1.0,valign=c]{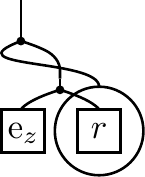}
    \hspace{0.06em}=
    \includegraphics[scale=1.0,valign=c]{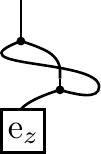}
    \hspace{0.06em}=-\hspace{0.08em}
    \includegraphics[scale=1.0,valign=c]{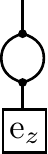}
    \hspace{0.06em}=\hspace{0.06em}
    \adjustbox{valign=c}{\begin{tikzpicture}[]
        \node (O) at (0,0) [squ,draw] {2};
    \end{tikzpicture}}
    \hspace{0.06em}
    \includegraphics[scale=1.0,valign=c]{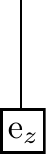}
\end{equation}
The last step is due to $\,\includegraphics[scale=1.0,valign=c]{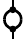}=-\hspace{0.2em}
\adjustbox{valign=c,scale=0.8}{\begin{tikzpicture}
    \node (yee) at (0,0) [squ,draw] {$2$};
\end{tikzpicture}}\,
\includegraphics[scale=1.0,valign=c]{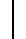}\hspace{0.17em}$, which can be proved by the following.
\begin{equation}
    \label{eq:vacpolarization}
    \adjustbox{valign=c}{\includegraphics[valign=c,scale=1,angle=90]{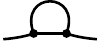}}
    \,\,\,\,=\,\,\,\,
    \adjustbox{valign=c}{\includegraphics[valign=c,scale=1,angle=90]{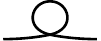}}
    \hspace{0.1em}-\hspace{0.1em}
    \adjustbox{valign=c}{\includegraphics[valign=c,scale=1]{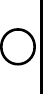}}
    \,\,\,\,=\,\,
    \,\hspace{-0.15em}
    \hspace{-2mm}
    \adjustbox{valign=c}{\includegraphics[valign=c,scale=1,angle=90]{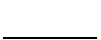}}\,\,\,\,
    -
    \,\,\,
    \adjustbox{valign=c}{\begin{tikzpicture}
        \node (O) at (0,0) [squ,draw] {$3$};
    \end{tikzpicture}}
    \hspace{-2mm}
    \adjustbox{valign=c}{\includegraphics[valign=c,scale=1,angle=90]{figs/APPI.pdf}}\,
    \,\,\,\,=\,\,\,\,
    -
    \,\,
    \adjustbox{valign=c}{\begin{tikzpicture}
        \node (O) at (0,0) [squ,draw] {$2$};
    \end{tikzpicture}}
    \hspace{-2mm}
    \adjustbox{valign=c}{\includegraphics[valign=c,scale=1,angle=90]{figs/APPI.pdf}}
\end{equation}
Rather than using coordinate expressions of gradient, curl, and divergence in particular coordinates, working in a coordinate-free manner has several benefits. In complex cases, it can be faster and has a lower possibility to make mistakes.\cite{Note9} Also, it offers an algebraic way to find the $\delta^{(3)}(\vec{r})$ term in the divergence or curl of a vector field.\cite{SuppMat} It is notable that such advantages are doubled with the graphical notation that significantly lowers the difficulty of handling higher-rank index manipulations. For various physical examples such as dipolar electromagnetic fields and flow configurations in fluid dynamics, refer to the supplementary material.\cite{SuppMat}

%There are also various physical examples\cite{SuppMat} such as dipolar electromagnetic fields and flow configurations in fluid dynamics. While working on those exercises, students would find that many of them stems from few common identities those which are sometimes tensorial. (Despite the fact that tensors are not explicitly introduced by the educator!) The graphical notation will clearly reveal such shared structures. This will guide them to figure out what vector field configurations are geometrically of a same footing (e.g. of a same multipole order).

\subsection{A First Look on Tensors}
Lastly, we want to comment about tensors, since they occasionally appear in undergraduate physics. Students are likely to develop the ideas of tensors by themselves while utilizing the graphical vector calculus; the extension from zero and one-terminal objects to multi-terminal objects is straightforward, and the graphical notation naturally involves the manipulation of multiple terminals. Also, graphical representations are useful to explain the concept of tensors to students, utilizing the ``machine view.'' For example, think about the inertia tensor,
$
I_{ij}
=
\adjustbox{valign=c, scale=0.8}{\begin{tikzpicture}[]
    \node (O) at (0,0) [squ,draw] {$I$};
    \node (i) at (-0.8,0) [empty, inner sep=1pt] {\imarker{$i$}};
    \node (j) at (+0.8,0) [empty, inner sep=1pt] {\imarker{$j$}};
    \draw (i)--(O)--(j);
\end{tikzpicture}}
$. It is simply a two-terminal device that ``modulates'' a one-terminal input (angular velocity,
$\adjustbox{scale=0.8, valign=c}{\begin{tikzpicture}
    \node (A) at (0,0) [squ, draw] {$\omega$};
    \draw (A) -- (-0.75,0);
\end{tikzpicture}}$
) into a one-terminal output (angular momentum,
$\adjustbox{valign=c,scale=0.8}{\begin{tikzpicture}
    \node (A) at (0.75,0) [squ, draw] {$L$};
    \node (O) at (0,0) [empty] {};
    \draw (A) -- (O);
\end{tikzpicture}}
=
\adjustbox{valign=c,scale=0.8}{\begin{tikzpicture}
    \node (O) at (0.6,0) [squ,draw] {$I$};
    \node (i) at (0,0) [empty] {};
    \node (j) at (1.2,0) [squ,draw] {$\omega$};
    \draw (i)--(O)--(j);
\end{tikzpicture}}\,
$). Imagine as if a ``signal'' generated from the $\omega$ box propagates from right to left. Swapping the two arms of the inertia tensor does not affects the value), because it is symmetric:
$
\adjustbox{valign=c, scale=0.8}{\begin{tikzpicture}[]
    \node (O) at (0,0) [squ,draw] {$I$};
    \node (i) at (-0.8,0) [empty, inner sep=1pt] {\imarker{$i$}};
    \node (j) at (+0.8,0) [empty, inner sep=1pt] {\imarker{$j$}};
    \draw (i)--(O)--(j);
\end{tikzpicture}}
= I_{ij} = I_{ji} =
\adjustbox{valign=c, scale=0.8}{\begin{tikzpicture}[]
    \node (O) at (0,0.4) [squ,draw] {$I$};
    \node (cx) at (0.35,0) [empty] {};
    \node (cR) at ($(O)+(cx)$) [empty] {};
    \node (cL) at ($(O)-(cx)$) [empty] {};
    \node (sx) at (0.67,0) [empty] {};
    \node (sL) at ($(0,0)-(sx)$) [empty] {};
    \node (sR) at ($(0,0)+(sx)$) [empty] {};
    \node (i) at (-0.8,0) [empty, inner sep=1pt] {\imarker{$i$}};
    \node (j) at (+0.8,0) [empty, inner sep=1pt] {\imarker{$j$}};
    \draw (i)..controls(sR)and(cR)..(O);
    \draw (j)..controls(sL)and(cL)..(O);
\end{tikzpicture}}
=
\adjustbox{valign=c, scale=0.8}{\begin{tikzpicture}[]
    \node (O) at (0,0) [squ,draw] {$I$};
    \node (sy) at (0,0.3) [empty] {};
    \node (s) at ($(0.2,0)+(sy)$) [empty] {};
    \node (I) at (s) [empty] {};
    \node (J) at ($(O)-(s)$) [empty] {};
    \node (i) at ($(-0.8,0)+(sy)$) [empty, inner sep=1pt] {\imarker{$i$}};
    \node (j) at ($(0.8,0)-(sy)$) [empty, inner sep=1pt] {\imarker{$j$}};
    \draw (i)--(I) to[out=0,  in=0,  looseness=2] (O);
    \draw (j)--(J) to[out=180,in=180,looseness=2] (O);
\end{tikzpicture}}
$.
However, this is not the case for a general multi-terminal object unless it is symmetric, as we have already discussed in \cref{sec:meetthecrosspoduct}. For the details of graphical representations of such general objects, refer to the supplementary material.\cite{SuppMat} Here, we restrict our attention to symmetric rank-2 tensors.

At least there are three of the practical benefits of using graphical notation for tensor equations. First, it is convenient to calculate the trace\cite{Note10} and related quantities of a tensor.\cite{SuppMat} Next, the graphical notation provides a transparent and unambiguous way to denote contraction structures. For example, consider the two expressions below denoting $K=\frac{1}{2}\omega_{i}I_{ij}{\omega}_{j}$ and $\epsilon_{ijk}\omega_{j}L_k = \epsilon_{ijk}\omega_{j}I_{kl}\omega_{l}$ respectively,
\begin{equation}
    \adjustbox{valign=c}{\begin{tikzpicture}
        \node (O) at (0,0) [squ,draw] {$K$};
    \end{tikzpicture}}
    =
    \adjustbox{valign=c}{\begin{tikzpicture}
        \node (O) at (0,0) [squ,draw] {\adjustbox{valign=c,scale=0.75}{$\textstyle \frac{1}{2}$}};
    \end{tikzpicture}}
    \,\,
    \adjustbox{valign=c}{\begin{tikzpicture}
        \node (O) at (0,0) [squ,draw] {$I$};
        \node (A) at (-0.7,0) [squ,draw] {$\omega$};
        \node (B) at (0.7,0) [squ,draw] {$\omega$};
        \draw (A)--(O)--(B);
    \end{tikzpicture}}\,\,,
    \,\,
    \adjustbox{valign=c}{\begin{tikzpicture}
        \node (I) at (0,0) [empty] {};
        \node (Q) at (-0.2,0) [empty] {};
        \fill (Q) circle (1.3pt);
        \node (i) at ($(Q)+(0,0.8)$) {};
        \node (B) at (0.3,0) [squ,draw] {$L$};
        \node (C) at ($(Q)+(-120:0.5)$) [squ,draw] {$\omega$};
        \draw (i)--(Q)--(B);
        \draw (Q) to[out=-150,in=75] (C);
    \end{tikzpicture}}
    \,\,=
    \adjustbox{valign=c}{\begin{tikzpicture}
        \node (I) at (0,0) [squ,draw] {$I$};
        \node (Q) at (-0.5,0) [empty] {};
        \fill (Q) circle (1.3pt);
        \node (i) at ($(Q)+(0,0.8)$) {};
        \node (B) at (0.7,0) [squ,draw] {$\omega$};
        \node (C) at ($(Q)+(-120:0.5)$) [squ,draw] {$\omega$};
        \draw (i)--(Q)--(I)--(B);
        \draw (Q) to[out=-150,in=75] (C);
    \end{tikzpicture}}\,,
\end{equation}
or the following more complex example that appears in the formula for the angular profile of electric quadrupole radiation power.
\begin{equation}
    \Bigg[\,\hspace{0.05em}\includegraphics[valign=c]{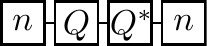}-\includegraphics[valign=c]{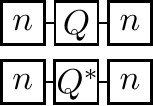}\,\,\Bigg]
\end{equation}
Here, $
\adjustbox{valign=c, scale=0.8}{\begin{tikzpicture}[]
    \node (O) at (0,0) [squ,draw] {$Q$};
    \node (i) at (-0.8,0) [empty, inner sep=1pt] {\imarker{$i$}};
    \node (j) at (+0.8,0) [empty, inner sep=1pt] {\imarker{$j$}};
    \draw (i)--(O)--(j);
\end{tikzpicture}}
= Q_{ij}$ is the electric quadrupole moment which is also a symmetric tensor. The asterisk stands for complex conjugation. For a calculus example, consider the divergence of the stress tensor $\boldsymbol{\sigma}$, $\nabla\cdot\boldsymbol{\sigma}$. Which index of $\boldsymbol{\sigma}$ is in charge of the inner product in the expression ``$\nabla\cdot\boldsymbol{\sigma}$?'---find the answer in the following diagram.
\begin{equation}
    \adjustbox{valign=c, scale=1}{\begin{tikzpicture}[]
    \node (i) at (-0.8,0) [empty] {};
    \node (j) at (+0.5,0) [empty] {};
    \node (z1) at ($(j)+(0.2,-0.2)$) [empty] {};
    \node (z2) at (0.25,-0.4) [empty] {};
    \draw [rounded corners=0.2cm] (j) -| (z1) |- (z2);
    \node (dif) at (O) [empty] {\includegraphics[scale=1,valign=c]{figs/CCSqBalloon9mm.pdf}};
    \node (O) at (0,0) [squ,draw] {$\sigma$};
    \draw (i)--(O)--(j);
\end{tikzpicture}}
\end{equation}
The contraction structures and their symmetry are clearly evident at a glance and can be quickly denoted in an unambiguous and less-bulky form, in comparison to ordinary notations.
Moreover, as one finds in the supplementary material,\cite{SuppMat} one can wisely calculate enormous tensor expressions in a shortcut with the guidance of the graphical notation. Lastly, the graphical notation is considerably useful in denoting and explaining the invariance property of tensorial expressions. As elaborated in the supplementary material,\cite{SuppMat} one can easily examine how the terminals of a tensor expression transforms under rotation intuitively by ``arrow pushing''---the pair creation/annihilation and propagation of arrowheads.

\section{Conclusions}
Graphical notations of tensor algebra have a history spanning over a century.\cite{cvitanovic2008grouptheory} The basic idea can be traced back to the late $19^{\text{th}}$ century works on invariant theory that related invariants to graphs.\cite{sylvester1878atomictheoryinvariants,clifford1878extract,kempe1885application,cayley1857xxviii} In the mid-$20^{\text{th}}$ century, diagrammatic methods such as Levinson and Yutsis' diagrams for $3n$-$j$ symbols\cite{levinson1956sumsofWignercoeff,yutsis1960mathematical} and Cvitanovi{\'{c}}'s birdtracks\cite{Cvitanovic1976grouptheoryforfeyn,cvitanovic1982spinorsinnegdim,cvitanovic2008grouptheory} are devised to conduct group-theoretic calculations and applied to quantum theory.\cite{canning1978diagrammatic,cvitanovic1981gauge,paldus1977application,stedman2009diagramtechniques} According to Levinson,\cite{levinson1956sumsofWignercoeff} one of the major motivations to develop such apparatus was ``the extreme inconvenience due to the bulkiness'' of the ordinary plaintext notation. On the other hand, Penrose\cite{penrose1957tensormethodsinag,penrose1971negativedimensionaltensors} devised a graphical notation for tensor algebra and utilized it in tensors and spinors in general relativity, theory of angular momentum and spin networks, and twistor theory.\cite{penroserindler1987spinorsandspace-time,penrose1973twistor,penrose2004roadtoreality} Similar to Levinson,\cite{levinson1956sumsofWignercoeff} one of his motivation was also to simplify the complicated equations and to effectively grasp the various interrelations they have by visual reasoning;\cite{penrose2010collectedvol1p25} however, he was also intended to introduce the concept of ``abstract tensor system'' by a coordinate-free notation that transparently retains the full syntactic structures of tensor equations.\cite{penrose1971negativedimensionaltensors,penroserindler1987spinorsandspace-time,penrose2004roadtoreality} The concept of the abstract tensor system and the Penrose graphical notation motivated the study of category theory and its graphical language in algebraic geometry,\cite{joyal1991geometryoftensorcalculus1,freyd1989braided,selinger2010survey,coeckeduncan2011interacting} and served as a background\cite{coeckeduncan2011interacting} to ``language engineering'' works to physics,\cite{coecke510032kindergarten,Coecke2011newstructuresforphysics,coecke2010categories} such as diagrams in tensor network of states\cite{selinger2004towards,AbC04categoricalsemantics,BB11categorical,biamonte2013tensornetworkmethodsforinvarianttheory,DBJC2012algebraicallytns,SinghVidal12tnsalgorithms} or quantum information and computing.\cite{coecke510032kindergarten,coeckeduncan2011interacting,coecke2017picturing}

So, why is the three-dimensional Euclidean vector calculus so quiet with such ``graphicalism?'' Perhaps it has been already being used as a private calculation technique, but its intractability to be printed due to graphical format might hindered its publication.\cite{penrose1971angularmomentum,joyal1991geometryoftensorcalculus1} However, regarding the popularity of Feynman diagrams that is also a graphical notation, it is worth casting light on the graphical tensor notation, as graphical vector calculus has its own pedagogical benefits. (Moreover, it conceptually precedes to Feynman diagrams.)
On the other hand, educators, already well-acquainted with the index notation and less sensible to beginners' difficulties, might have not tried to employ a graphical machinery to do vector calculus. However, there are introductory materials for graphical vector algebra and linear algebra,\cite{blinn2002quartic,blinn2002notationnotationnotation,peterson2009unshacklingLA,peterson2009onadiagrammaticproofofthecayley-hamiltontheorem,peterson2007notsocharacteristicequation} where differentiation does not comes into play. Therefore, publishing an educator's manual for the application of the graphical notation in vector calculus would be a useful thing to do.

What is newly proposed in this work is the graphical derivations and tricks of the vector differential calculus. No previous publications have dealt with the differentiation and integration of vector fields, while the graphical vector algebra introduced in this paper can be found also in other publications.\cite{peterson2009unshacklingLA,blinn2002quartic,keppeler2018birdtracks,stedman2009diagramtechniques,cvitanovic2008grouptheory} Also, pedagogical values of the graphical notation are demonstrated, and sufficient exercises containing both mathematical and physical calculations are provided. Overall, this paper will serve as a self-contained educational material.

The graphical notation has a lot of advantages. First, it provides a quick mnemonic or derivation for identities (e.g. Eq. \ref{eq:theidentity} or the vector calculus identities). It also enhances the calculation speed,\cite{omitboxes} giving a bird\textsc{\char13}s eye view to calculation scenario. The strategy of reducing complicated expressions can be wisely decided.
Although they are best performed in the graphical environment, such techniques on index gymnastics gained from graphical representations are inherited altogether into the index notation environment. An index notation user also will benefit from association of a tensorial expression with a graphical image.

Next, it has advantages in denoting and comprehending tensors.
If it is unambiguous, an index-free notation is preferred, that is, ``$\nabla \times \vec{A}$'' is preferred over ``${{\basise}_{i}}{{\mathsf{\epsilon }}_{ijk}}{{\partial }_{j}}{{A}_{k}}$,'' probably because it is more simple and easy to read off the tensorial structure in groups of semantic units (such as parsing $\vec{B}\cdot \nabla \times \vec{A}$ into ``$\vec{B}$ dot $\nabla \times \vec{A}$,'' not ``($\vec{B}$ cross $\nabla$) dot $\vec{A}$'').
Particularly, the graphical notation is preferable to other index-free notations, because it can flexibly represent tensor equations which become bulky in the ordinary index-free notation and transparently displays the contraction structure. The symmetry of a tensorial expression also can be grasped at a single glance.
Moreover, students will automatically discover the concept of tensors as an invariant $n$-terminal object and develop essential ideas of tensors in a coordinate-free setting using the graphical notation. For example, students will realize themselves interpreting the first term in the right hand side of Eq. \ref{eq:delAdotBLeibniz} as Eq. \ref{eq:delA} contracted with $\vec{B}$ at its second terminal (``input slot''). As a result, the idea of the tensor ``$\nabla\vec{A}$'' can be understood without leaving a vague impression, as its graphical representation provides a concrete comprehension of its functionality (as a ``machine'').
As parse trees (graphs) can promote understanding syntactic structures and generating sentences of the same structure, the graphical representation can do the same in tensor calculus and its education.\cite{Note11}
Furthermore, an unsupervised acquisition of tacit knowledge during graphical manipulation experiences such as ``the equations are also valid after undressing test vectors from them'' (\cref{sec:tripleproducts}) or ``a compound $n$-terminal object that has a permutation symmetry can be reduced into a simpler expression of the same symmetry up to a proportionality constant''\cite{SuppMat} is also notable.\cite{Note12}

Finally, it serves as an excellent primer to the graphical languages of advanced physics for undergraduates. After learning the graphical vector algebra, one can easily learn the birdtracks notation that is capable of group-theoretic calculations in quantum theory. Also, the graphical vector calculus provides exercises of ``diagrammatics,'' translating equations into graphics and vice versa that is an everyday task when one learns quantum field theory. Enthusiastic undergraduates who have always been curious about the working principles of Feynman diagrams will quench their thirst by learning the graphical tensor algebra. In essence, graphs for tensorial expressions of various symmetry groups, birdtracks, is a group-theoretic portion of Feynman diagrams. It is easy to learn Feynman diagrams after learning birdtracks or graphical tensor algebra and vice versa because the way they denote mathematical structures is alike: loop diagrams for trace (``vacuum bubbles,'' Eq. \ref{eq:vacbubble}) or etc. Meanwhile, birdtracks may leave a more concrete impression because it has graphical ``progression rules''\cite{Note13} that enables to jump from an expression to another via equality unlike Feynman diagrams. Furthermore, when one considers a series expansion of a tensorial expression, one encounters the exact parallel with diagrammatic perturbation in statistical mechanics or quantum field theory. Pedagogical examples can be found in the supplementary material.\cite{SuppMat}

The core characteristic that provides a background to all these advantages is the ``physically implemented syntax'' of the notation. It is believed that Feynman diagrams work because it is indeed a faithful representation of the physical reality (to the best of our knowledge)---the nature is implemented by worldlines of particles that are isomorphic to Feynman diagrams.
In the graphical notation of tensors, the grammar of tensors is ``embodied'' in the wires, 3-junctions, nodes, beads, and all that: the symbols behave as its physical appearance (\textit{self-explanatory design} of symbols in \cref{sec:meettheKroneckerdelta} and \cref{sec:meetthecrosspoduct}).
Consequently, the language is highly intuitive and automatically simplifies tensorial expressions (\textit{the economy of the graphical notation}).
The association of a kinesthetic imagery further simplifies the perception and manipulation of the elements (\textit{the dancing rule} and the ``clank'' in \cref{sec:meetthecrosspoduct}).
As Feynman diagrams are the most natural language to describe the microscopic process of elementary particles, the graphical notation is the canonical language of the vector calculus system.

Last but not least, the graphical notation will change a vector calculus class into an enjoyable game.
As a child playing with educational toys such as Lego blocks or magnetic building sticks, it will be an entertaining experience to ``doodle'' with the dancing diagrams.
Even a calculation of complicated tensorial invariants can be a challenging task that thrills a person; one would feel as if he or she is doing cat\textsc{\char13}s cradle or literally ``gymnastics'' involving their visual, kinesthetic, or even multimodal neural substrates.
Such an amusing character can attract students’ interest and offer a motivation to study vector calculus.
Students would voluntarily build various tensorial structures, heuristically find the identities, and gain intuitions.
One possible ``creative classroom'' scenario can be suggested is to present students only the basic grammar of the graphical notation and letting them spontaneously and exploratively find the ``sentences (identities),'' perhaps in a group. The teacher can collect their results and have a group presentation, then introduce missing identities if any. This will turn a formula-memorizing class into an amusing voluntary learning experience.
So, how about boosting your education by the graphical notation?

\section{Acknowledgement}
We thank Elisha Peterson for the provision of access to his research materials and clarification on the reason why he gave the name ``trace diagrams'' to his diagrams via e-mail. The work of K.-Y. Kim was supported by Basic Science Research Program through the National Research Foundation of Korea (NRF) funded by the Ministry of Science, ICT \& Future Planning (NRF2017R1A2B4004810) and GIST Research Institute (GRI) grant funded by the GIST in 2019.
}

% Bibliography


\begin{thebibliography}{65}
\bibitem [{Note100()}]{omitboxes}%
  \BibitemOpen
  \bibinfo {note} {Tip: during quick calculations, you can omit surrounding characters with boxes.}\BibitemShut {Stop}%
\end{thebibliography}
\end{document}